\documentclass[pra,twocolumn,10pt,aps,showpacs,showkeys]{revtex4-1}
\usepackage[english]{babel}
\usepackage[intlimits]{amsmath}
\usepackage{verbatim}
\usepackage{dcolumn}
\usepackage[utf8]{inputenc}
\usepackage[T1]{fontenc}
\usepackage[english]{babel}
\usepackage{epsfig}
\usepackage{lipsum}
\usepackage{graphicx,color}
\usepackage{amssymb}
\usepackage{mathtools}
\usepackage{dsfont}
\usepackage{stackrel}
\usepackage{calligra}
\usepackage{mathrsfs}
\usepackage[cal=boondoxo]{mathalfa}
\usepackage{hyperref}
\usepackage{float}

\newcommand{\eq}[1]{Eq.\;(\ref{#1})}

\makeatletter
\let\@hbar\hbar
\def\math@bold{bold}
\def\hbar{\ifx\math@version\math@bold\hbar@fett\else\@hbar\fi}
\def\hbar@fett{\mathchar'26\mkern-9muh}
\makeatother

 \newcommand{\bra}[1]{\langle{#1}|}
 \newcommand{\ket}[1]{|{#1}\rangle}

\renewcommand{\Re}{\operatorname{Re}}
\renewcommand{\Im}{\operatorname{Im}}

\usepackage{ifthen} 
\newcommand{\bu}[1][]{
	\ifthenelse{\equal{#1}{}}
	{\vec{u}}
	{\vec{u}_{#1}}
}
\newcommand{\bv}[1][]{
	\ifthenelse{\equal{#1}{}}
 	{\vec{v}}
	{\vec{v}_{#1}}
}

\def\vec#1{\ensuremath{\mathchoice{\mbox{\boldmath$\displaystyle#1$}}
                            {\mbox{\boldmath$\textstyle#1$}}
                            {\mbox{\boldmath$\scriptstyle#1$}}
                            {\mbox{\boldmath$\scriptscriptstyle#1$}}}}
\newcommand{\btu}[1][]{
	\ifthenelse{\equal{#1}{}}
	{\tilde{\vec{u}}}	
	{\tilde{\vec{u}}_{#1}}
}
 
\newcommand{\btv}[1][]{
	\ifthenelse{\equal{#1}{}}
	{\tilde{\vec{v}}}
	{\tilde{\vec{v}}_{#1}}
}

 \newcommand{\mycaption}[2]{\caption[#1]{#1\ifx\end#2\end\else . #2\fi}} 

\usepackage{textcomp}
\usepackage[squaren,cdot]{SIunits}

\begin{document}

\title{Two-body loss rates for reactive collisions of cold atoms}
\author{C. Cop}
\email{christiancop@online.de}
\author{R. Walser}
\affiliation{TU Darmstadt, Institut f\"ur Angewandte Physik, D-64289 Darmstadt, Germany}
\date{\today}

\begin{abstract}
We present an effective two-channel model for reactive collisions of cold 
atoms. It augments elastic molecular channels with an irreversible, inelastic loss channel. Scattering is studied with the  distorted-wave Born approximation and yields general expressions for
angular momentum resolved cross sections as well as two-body loss rates. 
Explicit expressions are obtained for piece-wise constant potentials. A pole expansion reveals simple 
universal shape functions for cross sections and two-body loss rates in agreement with 
the Wigner threshold laws. 
This is applied to collisions of metastable 
${^{20}}$Ne -- ${^{21}}$Ne atoms, which decay primarily through 
exothermic Penning-, or Associative- ionization processes. 
From a numerical solution of the multichannel Schrödinger equation using the best currently available molecular potentials, we have obtained synthetic scattering data.
Using the two-body loss shape functions derived in this paper, we can match these scattering data very well. 
\end{abstract}

\maketitle

\section{Introduction}
Understanding reactive collisions of metastable rare gas atoms 
(Rg*) is a central topic in trapping such species. Major loss 
processes in Rg* collisions are Penning Ionization (PI) and Associative 
Ionization (AI)~\cite{Penning1927,SISKA1993}
\begin{equation}
\label{piai}
\begin{aligned}
 \text{PI}:\quad \text{Rg*}\ +\ \text{Rg*} & \longrightarrow \text{Rg} \ + \ 
\text{Rg}^+ \ + \ e^-,\\
 \text{AI}:\quad \text{Rg*}\ +\ \text{Rg*} & \longrightarrow \text{Rg}_2^+ \ 
+\ e^-.
\end{aligned}
\end{equation}
In current trapping experiments~\cite{VASSEN2012}, it is possible to observe the reaction 
kinetic 
with high resolution and in real-time, as the ionic fragments can be detected with 
single-ion precision.

In order to parametrize cold ionizing collisions such as PI and AI, different models have been presented. The quantum reflection model~\cite{Friedrich2013} has been applied successfully to explain two-body loss 
rates 
in cold collisions of metastable rare gas atoms, for He* collisions~\cite{STAS2006,McNamara2007,DICKINSON2007}, Xe* collisions~\cite{Orzel1999} 
and 
Kr* collisions~\cite{Katori1995}. This model assumes complete ionization at short-range and predicts universal scattering rates for collisions of different isotope mixtures in agreement with mass scaling.

Current cold Ne* experiments of G. Birkl 
{\em et al.}~\cite{Spoden2005,Schutz2012,SCHUETZ2016} have provided new data on elastic as well as inelastic 
scattering. The isotope composition of  the Ne gas consists of the bosonic (B
) and fermionic (F) isotopes
with natural abundance, i.e.
$^{20}$Ne (90.48\%, B), $^{21}$Ne (0.27\%, F), 
$^{22}$Ne (9.25\%, B)~\cite{Wieser2009}. 
Homonuclear~\cite{Spoden2005,SCHUETZ2016} and heteronuclear 
collision rates~\cite{Schutz2012} were obtained, preparing polarized as well 
as 
unpolarized subensembles with respect to the internal magnetic substates. 
The experiment uncovers deviations from a fundamental mass scaling law for the 
elastic and inelastic scattering rates of different isotopes. Thus, a simple 
quantum reflection model is insufficient. 
\begin{figure}[H]
 \centering
 \includegraphics[clip,width=\columnwidth]{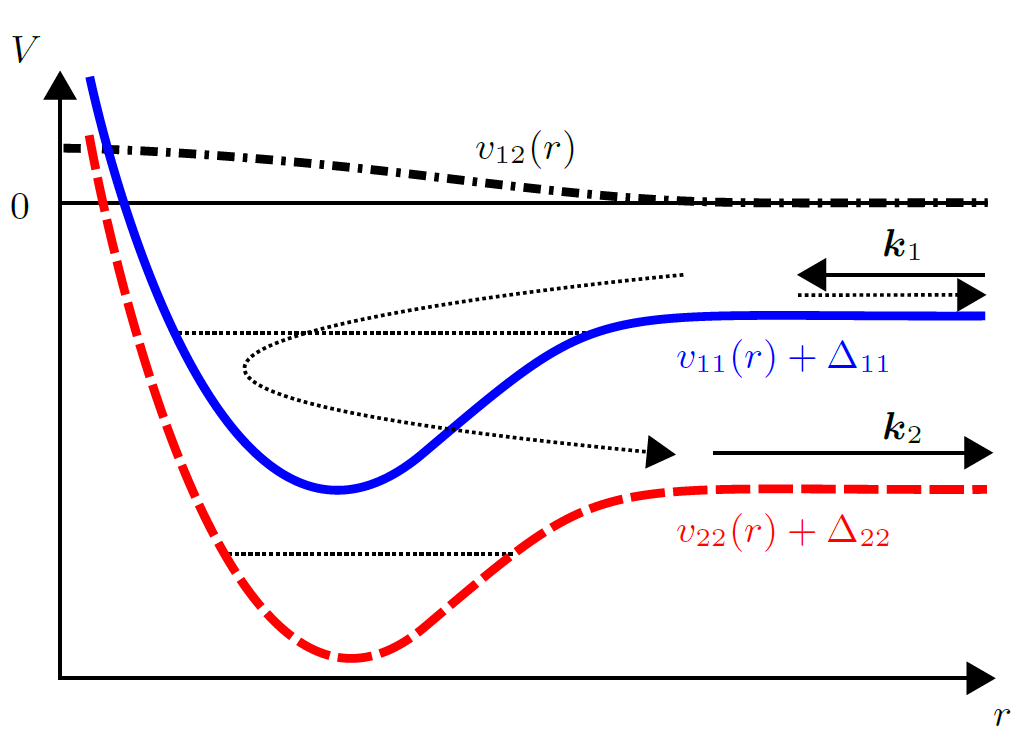}
 \caption{(Color online) Effective two-channel reaction model: 
molecular scattering channel 1 coupled to loss channel 2. 
Radial molecular 
potentials 
$v_{11}(r)$ (solid, blue) and loss potential $v_{22}(r)$ (dashed, red) 
are coupled by $v_{12}(r)$ (dash-dotted, black). 
At large separations, the potentials approach the 
channel 
thresholds $\Delta_{ii}$ with $\Delta_{11}>\Delta_{22}$ and all couplings $v_{ij}(r)$ vanish. 
The straight thin dashed black lines indicate bound states.
}
  \label{fig:twoch}
\end{figure}

To match these experimental facts, we consider an effective two-channel model using molecular 
potentials~\cite{Kotochigova2000,Kotochigova2015} with van-der-Waals interaction at long-range~\cite{Derevianko2000} for the elastic channels as shown in 
Fig.~\ref{fig:twoch}. The free parameters of the fictitious inelastic loss potential, as well as the coupling strength were obtained from a finite temperature fit to the scattering data with very good agreement. 
The threshold scattering rates of the numerical two-channel model are also in very good agreement with recent non-universal models for reactive collisions~\cite{IDZIASZEK2010,Jachymski2013,Jachymski2014}, which have been applied successfully to atom-ion collisions~\cite{Idziaszek2011}, polar molecule collisions~\cite{Julienne2009,QUEMENER2010,Idziaszek2010a,Kotochigova2010,Julienne2011} and collisions of other species~\cite{Frye2015}.
For details, we refer to Ref.~\cite{COP2016}.

In this article, we will complement the numerical approach
with an analytical study of the two-channel model using
piecewise constant couplings and potentials, only. Within this 
hypothesis, scattering can be studied analytically 
in the complex $k$- or $E$-plane. 
The coupling strength between the molecular channel and the loss channel 
should be weak on physical grounds. Therefore, we will employ the
distorted-wave Born approximation (DWBA), which is introduced in 
Sec.~\ref{sec:scattering_theory} and yields a general expression for the 
partial-wave resolved two-body loss rates. Moreover, we introduce the concepts of the pole expansion of the $S$-matrix. 
Coupled square-wells are an illuminating application of the previous discussion. Exact and DBWA solutions are presented in Sec.~\ref{sec:square_wells}. However, even this simple model yields explicit expressions, which are too complex for interpretation. 
Therefore, we will step back to single channel scattering in 
Sec.~\ref{sec:single_channel_squarewell} and revisit the well known results 
$s$-, $p$-, and $d$-wave scattering phases and cross sections \cite{WA2010}. Already there, the complex analytical expression conceal the essential physics, which is uncovered only by a pole expansion. 
In Sec.~\ref{sec:two_body}, we merge the insights from the previous sections and study the two-body loss rates. Exact square-well results for $s$-, $p$-, and $d$-waves are compared to explicit expressions (shape functions) using the pole approximation. These shape functions are consistent with the Wigner threshold laws and are depicted for some suitably chosen parameters.
Finally, in Sec.~\ref{sec:effective_theory}, we demonstrate the utility of these physically motivated shape functions. While originally derived for square well potentials, they are also very well suited to interpolate synthetic scattering data obtained from the 
two-channel van-der-Waals model for the $^{20}$Ne-$^{21}$Ne 
collisions~\cite{COP2016}. 
The article ends with conclusions and  two appendices regarding definitions of Riccati-Bessel functions and details on the pole expansion in the complex plane.

\section{Multichannel scattering }
\label{sec:scattering_theory}

In order to understand the reactive quantum kinetics of 
Eqs.~\eqref{piai} quantitatively, 
one needs to consider two coupled channels, as depicted in 
Fig.~\ref{fig:twoch}. There, a molecular state manifold or channel 1 can 
either scatter elastically within the manifold, or inelastically to a loss 
channel 2. This removes particles irreversibly from the interaction 
zone.  
In the rest-frame of the collision partners, the state manifold 
$\mathcal{H}=
\text{span}
\{\vec{k}\in \mathds{R}^3,i\in\{1,2\}| \ket{\vec{k},i}\}$
is spanned by plane-waves decorated with a channel subscript.
The energy 
\begin{align}
\label{sg}
	H &=H_0+V(r).
\end{align}
consists of a asymptotically free Hamiltonian operator $H_0$ and a short 
range, matrix-valued potential $V(r)$. To keep the notation compact, we will use natural units $\hbar=2 \mu=1$, with the reduced mass of the collision partners $\mu=m_1 m_2/(m_1+m_2)$. Thus, $H_0=\vec{p}^2+
\Delta$ denotes the relative kinetic energy of the collision partners with 
respect to the collision thresholds energies $\Delta_{ij}=\delta_{ij
}\Delta_{ii}$ with $\Delta_{11}>\Delta_{22}$.
Plane-waves are eigenfunctions  of the non-interacting two-particle system
\begin{align}
\label{fsg}
H_0\ket{\vec{k}_i,i} &=E\ket{\vec{k}_i,i}, &E=k_i^2+\Delta_{ii},
\end{align}
with channel wave-numbers $k_i(E)=\sqrt{E-\Delta_{ii}}$ determined by the energy $E$
 relative to the threshold energies 
$\Delta_{ii}$.
The short range molecular potential 
\begin{equation}
	V = V^\text{I} + V^\text{II} = 
	\begin{pmatrix} v_{11} & 0 \\ 0 & v_{22} 
\end{pmatrix} + 
\begin{pmatrix} 
0 & v_{12}\\ 
v_{12}^\ast & 0 \end{pmatrix}
	\label{eq:pot_decomposition}
\end{equation}
can be decomposed into the 
potentials  $V^\text{I}$ for each individual channel and their coupling 
$V^\text{II}$. 

\subsection{Two-potential formula for the T-matrix}
Scattering in the presence of two potentials can be described 
using the 
two-potential formula~\cite{TAY2006} for the $T$-matrix elements 
\begin{gather}
\label{eq:two_pot_formula}
	t_{ij}(\vec{k}_i,\vec{k}_j)= 
	\bra{\vec{k}_i,i}T \ket{\vec{k}_j,j}\\	
	=\bra{\vec{k}_i,i}V^\text{I} 
	\ket{\vec{k}_j,j,+^\text{I} }
	+ \bra{\vec{k}_i,i,-^\text{I} } V^\text{II} \ket{\vec{k}_j,j,+}\notag.
\end{gather}
This expression introduces the scattering states 
$\ket{\vec{k}_i,i,\pm^\text{I} }$ 
that are obtained in the absence of potential $V^\text{II}$ and the
fully coupled scattering states $\ket{\vec{k}_i,i,\pm}$. They 
are defined by their corresponding Lippmann-Schwinger equations as
\begin{align}
	\label{eq:LS_uncoupled}
	\ket{\vec{k}_i,i,\pm^\text{I} } &= \ket{\vec{k}_i,i} + 
	G^\pm_0 (E)_{ii} v_{ii}^\text{I} 
	\ket{\vec{k}_i,i,\pm^\text{I}},\\
	\label{eq:LS_full}
	\ket{\vec{k}_i,i,\pm} &= \ket{\vec{k}_i,i} + 
	\sum_{j=1}^2 G^\pm_0(E)_{ii} v_{ij} \ket{\vec{k}_j,j,\pm},
\end{align}
with
$G^\pm_0(E)\equiv 
\lim_{\epsilon\rightarrow +0}{(E\pm i \epsilon-H_0)^{-1}}$ 
denoting the free advanced (-) and retarded (+) Green's functions. 

The exact result \eq{eq:two_pot_formula} is most useful if the second potential is weak and the Born series expansion is applicable. 
To first order, the distorted-wave Born approximation for the T-matrix reads
\begin{gather}
	\label{eq:DWBA}
	t_{ij}(\vec{k}_i,\vec{k}_j)=
	t^{(0)}_{ij}(\vec{k}_i,\vec{k}_j)
	+t^{(1)}_{ij}(\vec{k}_i,\vec{k}_j)
	+\mathcal{O}\left(\protect{V^{\text{II}}}^2\right) \\
	=\bra{\vec{k}_i,i}V^\text{I} 
	\ket{\vec{k}_j,j,+^\text{I} }
	+ \bra{\vec{k}_i,i,-^\text{I} } V^\text{II} 
	\ket{\vec{k}_j,j,+^\text{I} }+\ldots\notag
\end{gather} 
It is important to note that first order corrections due to the second 
potential only occur 
in the off-diagonal $t_{i\neq j}$ matrix element.

\subsection{Scattering states in terms of Jost functions}
To proceed in the analysis of reactive two-body scattering, we will assume 
rotational invariant interactions \cite{COP2016}.
This implies angular momentum 
conservation and suggests to use symmetry adapted, spherical coordinates 
$(r,\theta,\phi)$. 
Single channel scattering states with outgoing- or incoming 
asymptotics ($\pm$) are conventionally~\cite{TAY2006}
defined as
\begin{equation}
\langle \mathbf{x}\ket{E,l,m,\pm}=
i^l\left(\frac{2}{\pi \sqrt{E}}\right)^{1/2}\frac{\psi_{l}^\pm(E,r)}{r}Y_{lm}
(\theta,\phi),
\end{equation} 
with the reduced scalar wave-functions $\psi_{l}^\pm(E,r)$, 
the eigenvalues of the 
angular projection $m$ on the quantization axis $z$ and angular momentum $l$
. If this is generalized to 
multichannel 
scattering,
the Schr\"odinger \eq{sg} reads
\begin{equation}
	\left(K^2(E)+\partial_r^2 -  V(r)-\frac{l(l+1)}{r^2}\right) 
		\Psi^\pm_{l}(E,r) = 0,
	\label{eq:tc_radial_schroed_matrix}
\end{equation}
where the wave-number matrix $K_{ij}(E)=\delta_{ij}k_i(E)$ is defined by the 
individual channel wave-numbers $k_i(E)$ as in \eq{fsg}. In turn, the 
reduced scattering wave-function 
\begin{equation}
	\Psi^\pm_{l}(E,r)= 
	\begin{pmatrix} 
	\psi_{l,11}^\pm(E,r) & \psi_{l,12}^\pm(E,r) \\ 
	\psi_{l,21}^\pm(E,r) & \psi_{l,22}^\pm(E,r) 
	\end{pmatrix},
\end{equation}
becomes a $2\times 2$ matrix to account for the two linearly independent 
solutions in the scattering channels. They are determined from
the boundary condition at the origin 	
\begin{align}
\Psi^\pm_{l}(E,r=0)=0
\end{align} 
and the asymptotic behavior for the outgoing solution
\begin{align}
	\Psi_{l}^+(E,r) \stackrel[r\to \infty]{}{\longrightarrow} & 
\frac{i}{2} \left\{ \hat{h}^-_l (Kr)- 
		\frac{\hat{h}^+_l(Kr)}{\sqrt{K}} S_{l}(E) \sqrt{K}\right\}, 
		\label{eq:psiplus_s}
\end{align}
as well as $\Psi_{l}^-=\left( \Psi_{l}^+\right)^\dag$.
Here, $\hat{h}_l^\pm$ are the Riccati-Hankel functions as defined in appendix 
\ref{sec:bessel} and $S_l(E)$ denotes the $S$-matrix with angular momentum $l$.

The {\em regular solution} $\Phi_{l}(E,r)$ is another solution to the Schr\"odinger equation \eq{eq:tc_radial_schroed_matrix}. It is defined by the boundary condition at the origin
\begin{equation}
	\Phi_{l}(E,r)\stackrel[r\to 0]{}{\longrightarrow} \hat{\jmath}_l(K r).
	\label{eq:regular_zero}
\end{equation}
In each channel, the solution approaches a Riccati-Bessel 
function $\hat{\jmath}_l(k_ir)$ close to the 
origin and is smaller in amplitude in all other coupled channels 
(cf. appendix \ref{sec:bessel}). 
Since the coupled radial 
Schr\"odinger equation \eq{eq:tc_radial_schroed_matrix} and the boundary 
condition \eq{eq:regular_zero} are real for real energies, the regular 
solution is real as well. In the asymptotic region the regular solution reads
\begin{equation}
	\Phi_{l}(E,r) \stackrel[r\to \infty]{}{\longrightarrow} \frac{i}{2} \left\{ 
\hat{h}^-_l (Kr) \mathcal{F}_{l}(K) - \hat{h}^+_l (Kr) \mathcal{F}_{l}(-K) 
\right\},
	\label{eq:regular_asymp}
\end{equation}
with the Jost matrix $\mathcal{F}_{l}(K)$ and its matrix elements 
$\mathcal{f}_{l,ij}$. Comparing the asymptotic form of the scattering wave-function 
\eq{eq:psiplus_s} to the asymptotic form of the regular solution 
\eq{eq:regular_asymp} leads to the relation of the scattering wave-function and 
the regular solution, given by
\begin{equation}
	\psi_{l}^+(E,r) = \Phi_{l}(E,r) \mathcal{F}_{l}^{-1}(K),
	\label{eq:phi_psi}
\end{equation}
and to the relation of the Jost matrix and the $S$-matrix
\begin{equation}
	S_{l}(E) = \sqrt{K}\mathcal{F}_{l}(-K) \mathcal{F}_{l}^{-1}/ 
	\sqrt{K}.
	\label{eq:Smatrix_Jost}
\end{equation}
Using the constitutive relation between the $S$ and the $T$ operator
\begin{equation}
	S (E)= \mathds{1} - 2\pi i \delta(E-H_0)T(E),
	\label{eq:S_T_relation}
\end{equation}
one can connect all relevant entities.
Within the distorted-wave Born approximation from \eq{eq:DWBA}, the $S$-matrix reads to first order as 
\begin{equation}
	S_{l}(E) = \begin{pmatrix} s_{l,11}^{(0)}(E) & s_{l,12}^{(1)}(E) \\ 
	s_{l,21}^{(1)}(E) & s_{l,22}^{(0)}(E) \end{pmatrix}
+\mathcal{O}\left(\protect{V^{\text{II}}}^2\right),
\end{equation}
where the diagonal $S$-matrix elements
\begin{equation}
	s_{l,ii}^{(0)}(E) = \frac{\mathcal{f}_{l,ii}^{(0)}(-k_i(E))}{
	\mathcal{f}_{l,ii}^{(0)}(k_i(E))}=e^{2 i \eta_{l,i}(k_i(E))}.
	\label{eq:s_jost_uncoupled}
\end{equation}
are the uncoupled solutions for the 
potential $V_\text{I}$ and given in terms of the uncoupled Jost functions 
$\mathcal{f}_{l,ii}^{(0)}(k_i)$, or equivalently by 
scattering phases $\eta_{l,i}(k_i)$. For the sake of readability, we have not included a further superscript $(0)$ in the scattering phase. The scattering phases
\begin{equation}
	\eta_{l,i}(k_i(E)) = \eta^b_{l,i} (k_i(E)) + \eta^r_{l,i} (k_i(E)),
	\label{eq:eta_decomp}
\end{equation}
can be decomposed into a background contribution $\eta^b_{l,i}$, which is a slowly varying function of $k_i$ and a resonant contribution $\eta^r_{l,i}$, which changes rapidly across a resonance. With this decomposition we can write for the $S$-matrix elements
\begin{equation}
	s_{l,ii}^{(0)}(E) = s_{l,ii}^{(0)b}(E)\cdot s_{l,ii}^{(0)r}(E).
\end{equation}

In the following, we prefer to switch from the energy parameter $E$ to the 
wave-number $k$. All functions will become functions of $k_i(E)$ while being 
implicitly functions of $E$ due to \eq{fsg}. 
The observable available for measurement is the partial elastic scattering 
cross section~\cite{TAY2006}  
\begin{equation}
	\sigma^{(0)}_{l,ii} (k_i) = 2\pi\frac{2l+1}{k_i^2}  
	(1-\Re s^{(0)}_{l,ii}(k_i) ).
	\label{eq:sigma_partial_elastic}
\end{equation}
The cross section for scattering of channel $1$ to channel $2$ in the DWBA is given by
\begin{equation}
	\sigma^{(1)}_{l,12} (k_1) = \pi\frac{2l+1}{k_1^2}  
	| s^{(1)}_{l,12}(k_1)|^2.
	\label{eq:sigma_partial_inelastic}
\end{equation}
The off-diagonal $S$ matrix-elements are obtained from the DWBA 
approximation \eq{eq:DWBA} and the constitutive relation~\eq{eq:S_T_relation} as
\begin{align}
	\label{eq:s12_DWBA}
	s^{(1)}_{l,12}(E) = & \frac{2 
	\int_0^\infty dr \ \psi^{(0)+}_{l,11} (E,r)  v_{12}^\text{II} (r) 
	\psi^{(0)+}_{l,22} (E,r)}{i\sqrt{k_1k_2}} \\
	= & e^{2i\eta_{l,12}(k_1(E))},\notag
\end{align}
where $\psi^{(0)+}_{l,ii} (E,r)$ are the uncoupled solutions of the 
radial Schr\"odinger \eq{eq:tc_radial_schroed_matrix} for the 
potential $V^\text{I}$ only, using 
$(\psi_{l,ii}^{(0)-}(E))^*=\psi_{l,ii}^{(0)+}(E)$.
The complex phase $\eta_{l,12}(k_1)\in\mathds{C}$ considers also attentuation 
and can be decomposed into a resonance- and a background contribution 
as in \eq{eq:eta_decomp}. 
 
The observable for measurement of atom loss in channel $1$ is the two-body loss rate~\cite{Orzel1999} which is given in terms of the cross section from channel $1$ to channel $2$ 
\begin{equation}
	\beta^{(1)}_l(k_1) = 2 k_1  \sigma^{(1)}_{l,12}(k_1)= 
	2\pi\frac{ 2l+1 }{k_1}|s_{l,12}^{(1)}(k_1)|^2.
	\label{eq:partial_twobody}
\end{equation}
It quantifies the loss of probability current from the elastic scattering 
channel 1.

\subsection{Pole expansion of the $S$-matrix}
\label{sec:pole_exp}

Scattering theory greatly benefits from complex analysis and the 
continuation of real parameters, like the energy $E$ or the wave-number $k$, 
into the complex plane.
For finite-range potentials, it can be shown that the Jost functions 
are entire functions of $k$. Then, the $S$-matrix  is
analytic everywhere in the complex $E$-, or $k$-plane except at singular 
points, when the Jost function vanishes~\cite{TAY2006}. 

For complex channel wave-numbers $k_i$, the analytic continuation of the 
uncoupled Jost functions reads
\begin{equation}
	\mathcal{f}_{l}(k)^* = \mathcal{f}_{l}(-k^*).
\end{equation}
With the help of the Weierstrass factorization theorem
\cite{Freitag2009,Conway2010,NEW2002}, one can
express the Jost function
\begin{align}
	\mathcal{f}_{l}(k) = & \mathcal{f}_{l}(0) e^{ik} 
	\prod_{n=1}^{N'} 
	\left( 1-i\frac{k}{\varkappa_n}\right)
	\prod_{n=1}^{N} 
	\left( 1+i\frac{k}{\mathcal{K}_n}\right) \notag \\
	& \times \prod_{n=1}^\infty 
	\left( 1+\frac{k}{\mathcal{k}_n^\ast}\right)
	\left( 1-\frac{k}{\mathcal{k}_n}\right),
	\label{eq:Jost_Weierstrass_decomp}
\end{align}
as an infinite  product of its zeros. 
For each of the $N'$  virtual states of the scattering potential, 
there is a zero of the Jost function on the negative imaginary axis at  
$k=-i\varkappa_n$ with $\varkappa_n>0$. For each of the $N$ bound states, the Jost function vanishes 
on the positive imaginary axis at $k=i\mathcal{K}_n$ with $\mathcal{K}_n>0$. 
Moreover, there are always infinitely many scattering resonances of the 
potential, which correspond to the zeros of the Jost function at 
$k=\mathcal{k}_n$ and $k=-\mathcal{k}_n^*$ with $\Re \mathcal{k}_n>0$. 

From this representation of the Jost function, one obtains the $S$-matrix of 
\eq{eq:s_jost_uncoupled} as an
 infinite product of poles in the complex $k=k(E)$ plane

\begin{align}
	s_{l}(E) = & e^{-2ik}
	\prod_{n=1}^{N'} 
	\frac{\varkappa_n+ik}{\varkappa_n-ik}
	\prod_{n=1}^N 
	\frac{\mathcal{K}_n-ik}{\mathcal{K}_n+ik}
	\notag
\prod_{n=1}^\infty 
	\frac{\mathcal{k}_n+k}{\mathcal{k}_n^\ast+k}
	\frac{\mathcal{k}_n^\ast-k}{\mathcal{k}_n-k}\\
	&=e^{2i \eta^{b}_l(k)} \, s^{r}_{l}(E).
	\label{eq:S_pole_exp}
\end{align}
For low-energy collisions, the analytical behavior of the $S$-matrix is 
dominated by the resonant contribution $s^r_l$ 
originating from states closes to the energy threshold.
The effect of the infinitely many other poles of the pole 
expansion can be summarized in a slowly varying background scattering phase 
$\eta^{b}_l$.

\section{Coupled square-wells}
\label{sec:square_wells}

\begin{figure}[H]
 \centering
 \includegraphics[clip,width=\columnwidth]{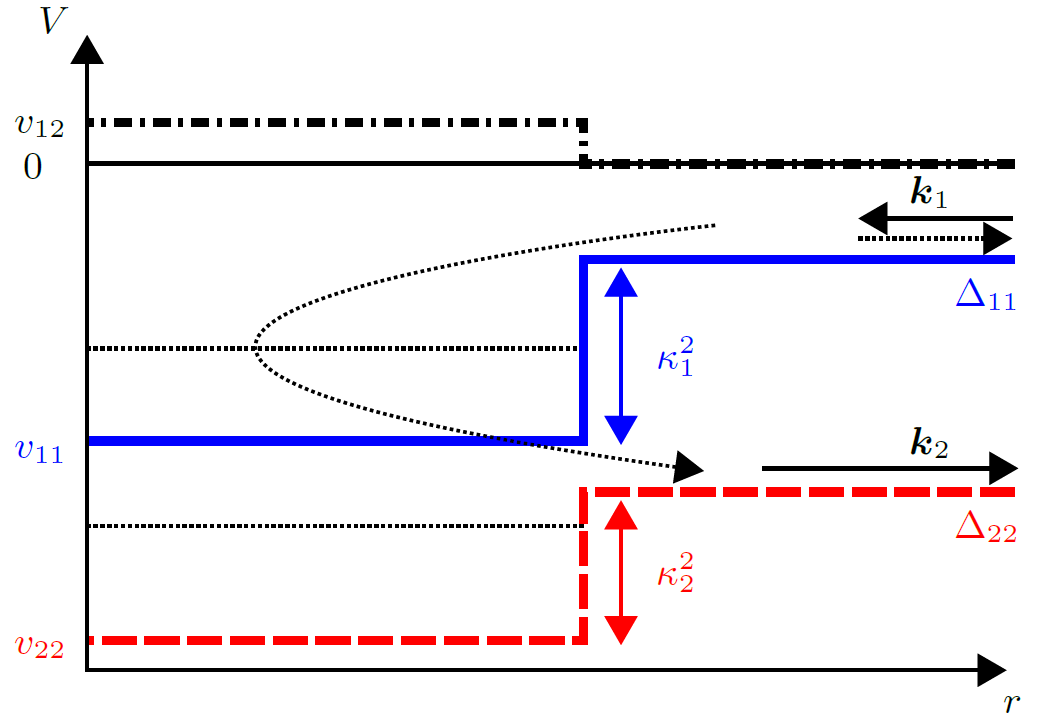}
 \caption{(Color online) Effective two-channel square-well potential vs. radius $r$. 
Within the range $r<1$, 
the potentials $v_{11}$ (solid, blue), $v_{22}$ (dashed, red) and $v_{12}$ (dash-dotted, black) are constant. For radii $r>1$, the potential matrix
decouples into the channel 
thresholds $\Delta_{11}$ and $\Delta_{22}$. The channel 
wave-numbers $k_1$, $k_2$ are defined as in \eq{fsg}. 
The straight thin dashed black lines indicate bound states.}
  \label{fig:box_twoch}
\end{figure}

In order to put these general considerations to practical use, 
we will study an elementary example of a weakly coupled 
square-well potential, 
which is shown in Fig.~\ref{fig:box_twoch}. 
There, the piecewise constant potential matrix reads
\begin{equation}
	v_{ij}(r) = \begin{cases} 
		v_{ij}& r\leq 1, \\ 
	\Delta_{ij} &r>1. \end{cases}
\end{equation}
The two levels $i\in \left\{ 1,2 \right\}$  are coupled within a
radius $r<1$ 
and decouple outside $\Delta_{i\neq j}=0$. 
The depths of the 
attractive potentials are parametrized by $\kappa_i$ and the inner level spacing is conveniently abbreviated by $\delta v$ as
\begin{align}
\kappa_i^2&=\Delta_{ii}-v_{ii}>0,\\
	\delta v&=v_{11}-v_{22}.
\end{align}

The utility of the model is seen by solving it 
perturbatively in Sec.~\ref{dwbasqwl} and exactly in Sec.~\ref{sec:exact_squarewell}.
The complexity of the solution conceals resonances
and a simple interpretation is possible only by the pole expansion presented in Sec.~\ref{sec:single_channel_squarewell}, eventually.

\subsection{Distorted-wave Born approximation}
\label{dwbasqwl}

First, to initialize the DWBA, we need the solution for the 
uncoupled square-well  Schr\"odinger 
\eq{eq:tc_radial_schroed_matrix} with $v_{12}=0$. 
In the inner and outer region the solution reads  
\begin{equation}
	\psi^{(0)+}_{l,ii}(E,r) = 
	\begin{cases} 
	(\frac{k_i}{q_i})^{l+1}
	\frac{\hat{\jmath}_l(q_i r)}{\mathcal{f}_{l,ii}^{(0)}(k_i)},
	& r\leq 1, \\ 
	\frac{s_{l,ii}^{(0)}(E) \hat{h}^+_l (k_i r)
	-\hat{h}^-_l(k_i r)}{2i} 
, &r>1, 
\end{cases}
	\label{eq:wf_squarewell}
\end{equation}
where we have introduced channel wave-numbers $q_i$ and $k_i$ for the inner 
and outer regions
\begin{align}
	q_1^2 &\equiv E - v_{11} =  k_1^2+\kappa_1^2, \\
	q_2^2 &\equiv E - v_{22} =  k_2^2+\kappa_2^2,
\end{align}
and $k_i$ as in \eq{fsg}.
The ansatz for the wave-function of 
\eq{eq:wf_squarewell} introduces the Jost function 
$\mathcal{f}^{(0)}_{l,ii}(k_i)$  by the relation between the regular solution 
at the  origin~\eq{eq:regular_zero} and the 
scattering solution \eq{eq:phi_psi}. 
By matching the wave-function smoothly at $r=1$, one obtains the Jost function 
$\mathcal{f}_{l,ii}^{(0)}(k_i) = g_l(k_i,q_i(k_i)),$
from the auxiliary function
\begin{equation}
\label{eq:f_ii_squarewell}
 g_l(k,q) = 2i \left[\frac{k}{q}\right]^{l} 
\frac{q \hat{\jmath}'_l(q) \hat{h}^+_l(k)-k \hat{h}_l^{+'}(k) 
	\hat{\jmath}_l(q)}{
	q (\hat{h}^{+'}_l(k) \hat{h}^-_l(k)
	-\hat{h}_l^{-'}(k) \hat{h}^+_l(k))},
\end{equation}
and the $S$-matrix elements from
\eq{eq:s_jost_uncoupled} as
\begin{align}
	s^{(0)}_{l,ii}(E) &=  
	\frac{
	q_i\hat{\jmath}_l'(q_i)\hat{h}^-_l (k_i)
	-k_i \hat{h}^{-'}_l(k_i) \hat{\jmath}_l(q_i)}{
	q_i\hat{\jmath}_l'(q_i) \hat{h}^+_l (k_i)
	-k_i \hat{h}^{+'}_l(k_i) \hat{\jmath}_l(q_i)}.
	\label{eq:s_ii_squarewell}
\end{align}

Second, in the next order of the DWBA the transition matrix element reads  
\begin{align}
	s_{l,12}^{(1)}(E) = & \frac{2 v_{12}  }{i \delta v \sqrt{k_1 k_2}}
	\left(\frac{k_1k_2}{q_1q_2}\right)^{l+1} 
	\frac{1}{\mathcal{f}^{(0)}_{l,11}(k_1) \mathcal{f}^{(0)}_{l,22} (k_2)}
	 \notag \\ 
	& \times [q_1\ \hat{\jmath}_{l-1} (q_1) \hat{\jmath}_l(q_2)-q_2\ 
	\hat{\jmath}_{l-1}(q_2)\hat{\jmath}_l(q_1)].
	\label{eq:s12_DWBA_squarewell}
\end{align}
using Eqs.~(\ref{eq:s12_DWBA}) and (\ref{eq:wf_squarewell}).
Thus, we obtain the interesting result that 
the transition matrix elements are given in terms 
of the uncoupled Jost functions. Thus, their zeros determine the poles  
of the $S$-matrix elements.

\subsection{Exact solution}
\label{sec:exact_squarewell}

Obtaining the full Jost matrix $\mathcal{F}_l$ for the coupled square-well 
analytically is a standard exercise. By matching the wave-function at the intersection, one finds for the diagonal elements 
\begin{align}
	\mathcal{f}_{l,11}(k_1) & = g_{l}(k_1,q_+)\cos^2 \alpha + g_{l}(k_1,q_-) 
\sin^2 \alpha, \\
	\mathcal{f}_{l,22}(k_2) & = g_{l}(k_2,q_-) \cos^2 \alpha + g_{l}(k_2,q_+) 
\sin^2 \alpha,	
\end{align}
and for the off-diagonal elements
\begin{align}
	\mathcal{f}_{l,12}(k_1) = & \frac{\sin 2\alpha}{2}  
	\left[\frac{k_2}{k_1} \right]^{l+1} \left[ g_{l}(k_1,q_+) - g_{l}(k_1,q_-) 
\right],
	\label{eq:f12} \\
	\mathcal{f}_{l,21}(k_2) = & \frac{\sin 2\alpha }{2} 
	\left[\frac{k_1}{k_2} \right]^{l+1} \left[ g_{l}(k_2,q_+) - g_{l}(k_2,q_-)
  \right],
\end{align}
defining a mixing angle $\alpha$ by $\tan 2 \alpha = 2 v_{12}/\delta v$.
The linearly independent solutions of the Schr\"odinger equation in the inner region are characterized by the two wave-numbers
$q_\pm =\sqrt{k^2-\varepsilon_\pm}$, with an energy splitting 
\begin{align}
	\varepsilon_\pm &= \frac{1}{2}(v_{11} + v_{22} 
\pm 
	\sqrt{\delta v^2+4 v_{12}^2}) .
\end{align}
Finally, the solution for the $S$-matrix can be found from the Jost matrix by \eq{eq:Smatrix_Jost}.
In particular, we are interested in the off-diagonal element and find
\begin{equation}
	s_{l,12}(E) = \sqrt{\frac{k_1}{k_2}}
	\frac{
	\mathcal{f}_{l,12}^*(k_1) \mathcal{f}_{l,11}(k_1)-\mathcal{f}^*_{l,11} (k_1)
	\mathcal{f}_{l,12}(k_1) }{\mathcal{f}_{l,11} (k_1)
	\mathcal{f}_{l,22}(k_2)-\mathcal{f}_{l,12}(k_1) \mathcal{f}_{l,21}(k_2)},
\end{equation}
with $k_i\equiv k_i(E)$. Clearly, one recovers the DWBA expression for $s_{l,12}^{(1)}$ of \eq{eq:s12_DWBA_squarewell}
from a first order expansion in  $\mathcal{O}(v_{12}/\delta v)$. 

\section{Single channel scattering}
\label{sec:single_channel_squarewell}

The understanding of the coupled channel scattering process is complex as 
single-channel scattering resonances are intertwined with two channel 
mixing. It is therefore prudent to dissect the problem and analyze 
single-channel scattering first. This is achieved by turning off the coupling  
$v_{12}=0$ between the channels and to study scattering in channel 1 exclusively. 
Now, we have the freedom to set the threshold energy $\Delta_{11}=0$. 

With the goal of parameterizing cross sections and loss rates
for cold collisions of atoms analytically,
we only consider the lowest $s$-, $p$-, $d$-
partial-waves. We will not exhaustively study all conceivable cases, 
but demonstrate the benefits of the pole expansion for certain instances.
In particular, for $s$-waves, we consider a potential
with a single weakly bound state in Sec.~\ref{subsubsec:sc_swave} 
and in Sec.~\ref{subsubsec:sc_pwave} for $p$- and $d$-waves with a single 
quasi-bound state.
These states dominate the behavior of the cross sections.
We will compare exact square-well results 
to the results from a single-pole expansion due to the weakly 
bound state for $s$-waves and from a two-pole expansion due 
to the quasi-bound states for $p$- and $d$-waves, eventually.

To enhance the visibility of resonances in the examples, we have 
deliberately chosen different potential depths for each angular momentum 
channel as listed in Table~\ref{tab:param_single}. 

\begin{table}[th]
\centering
\begin{ruledtabular}
\begin{tabular}{lcccc}
Angular momentum & $l $& $0$ (s)& $1$ (p)& $2$ (d)\\
\hline
Potential depth & $\kappa_1^2$ & $4.0$ & $9.0$ & $19.5$ \\
\hline
Zero of Jost function &$k_1$& $0.638i$ & $0.539$  & $0.628$  \\
& & & $-0.100i$ & $-0.009i$ \\
\hline
Binding/resonance  energy & $E_r$ & $-0.407$ & $0.281$ & $0.395$ \\
\end{tabular}
\end{ruledtabular}
\caption{Potential depth $\kappa_1^2$, complex zero $k_1$ of Jost function 
$\mathcal{f}^{(0)}_{l,11}(k_1)$ and
binding or resonance energy $E_r=\Re^2 k_1 - \Im^2 k_1$ in the uncoupled channel 1 for 
different angular momenta. Potential depths were deliberately chosen 
individually, so that each potential only supports one bound state ($s$), or 
one quasi-bound state ($p$,$d$).}
\label{tab:param_single}
\end{table}

\begin{figure}[H]
\centering\includegraphics[clip,width=\columnwidth]{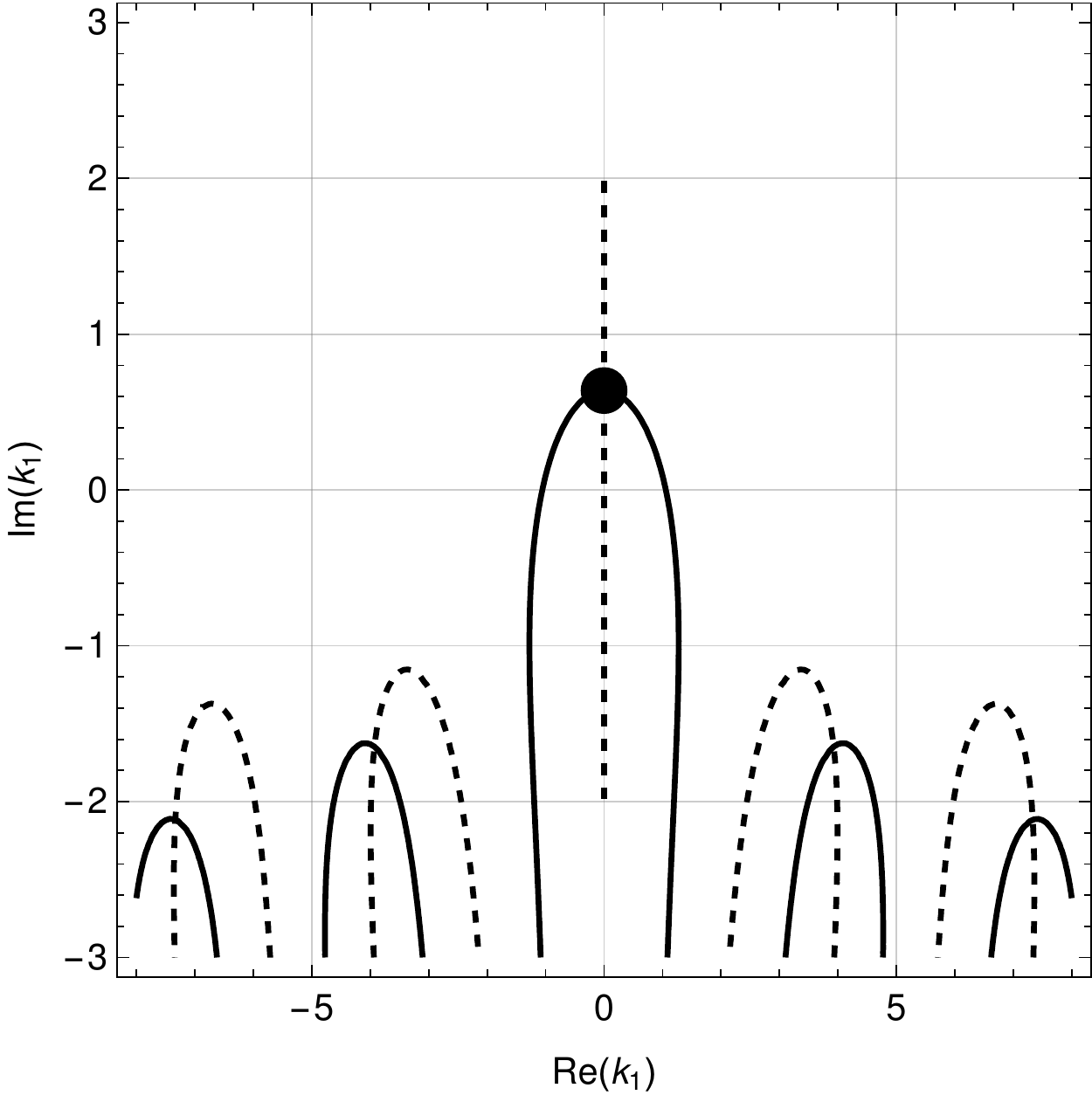}
\caption{Zero contour lines of the $s$-wave Jost function 
$\Re \mathcal{f}^{(0)}_{0,11}(k_1) = 0$ 
(solid) and 
$\Im \mathcal{f}^{(0)}_{0,11}(k_1) = 0$ (dashed) in the complex $k_1$-plane 
for a potential depth $\kappa_1^2=4$. The black dot indicates the position of a bound state at $k_1=i\mathcal{K}_1=0.638i$ and corresponds to a 
binding energy of $E_r=-\mathcal{K}_1^2=-0.407$.}
\label{fig:Jost_l0}
\end{figure}

\subsection{Scattering in the $s$-channel}
\label{subsubsec:sc_swave}

Shallow three-dimensional attractive square-wells can have zero, one, or
 more bound states. By increasing the potential depth continuously, a virtual 
state with positive energy transforms into a half bound state at zero energy 
and becomes a bound state with negative energy, eventually. In the following, 
we will present the exact scattering phase and cross section for $s$-waves and 
compare these with the pole expansion approximations.

\paragraph{Exact solution}

For $l=0$, the Jost function of \eq{eq:f_ii_squarewell} reads
\begin{equation}
	\mathcal{f}^{(0)}_{0,11}(k_1) = 
	\frac{e^{ik_1}  \sin q_1 }{q_1} ( q_1 \cot q_1 - i k_1 ),
	\label{eq:Jost_l0}
\end{equation}
and the $S$-matrix element follows from \eq{eq:s_jost_uncoupled} as
\begin{equation}
	s^{(0)}_{0,11}(k_1) = e^{-2ik_1} \frac{q_1 \cot q_1 + ik_1}{q_1 \cot q_1-ik_1} .
	\label{eq:sl0}
\end{equation}
For real $k_1$, it is unimodular $|s^{(0)}_{0,11}|=1$ and the real scattering phase reads
\begin{equation}
	\eta_{0}(k_1) =n\pi -k_1 + \arctan \frac{k_1}{q_1 \cot q_1} ,
\end{equation}
where Levinson theorem~\cite{TAY2006} determines the zero energy phase 
from the number of bound states $n$ in the scattering potential.

\paragraph{Single-pole expansion}

\begin{figure}[ht]
\centering\includegraphics[clip,width=\columnwidth]{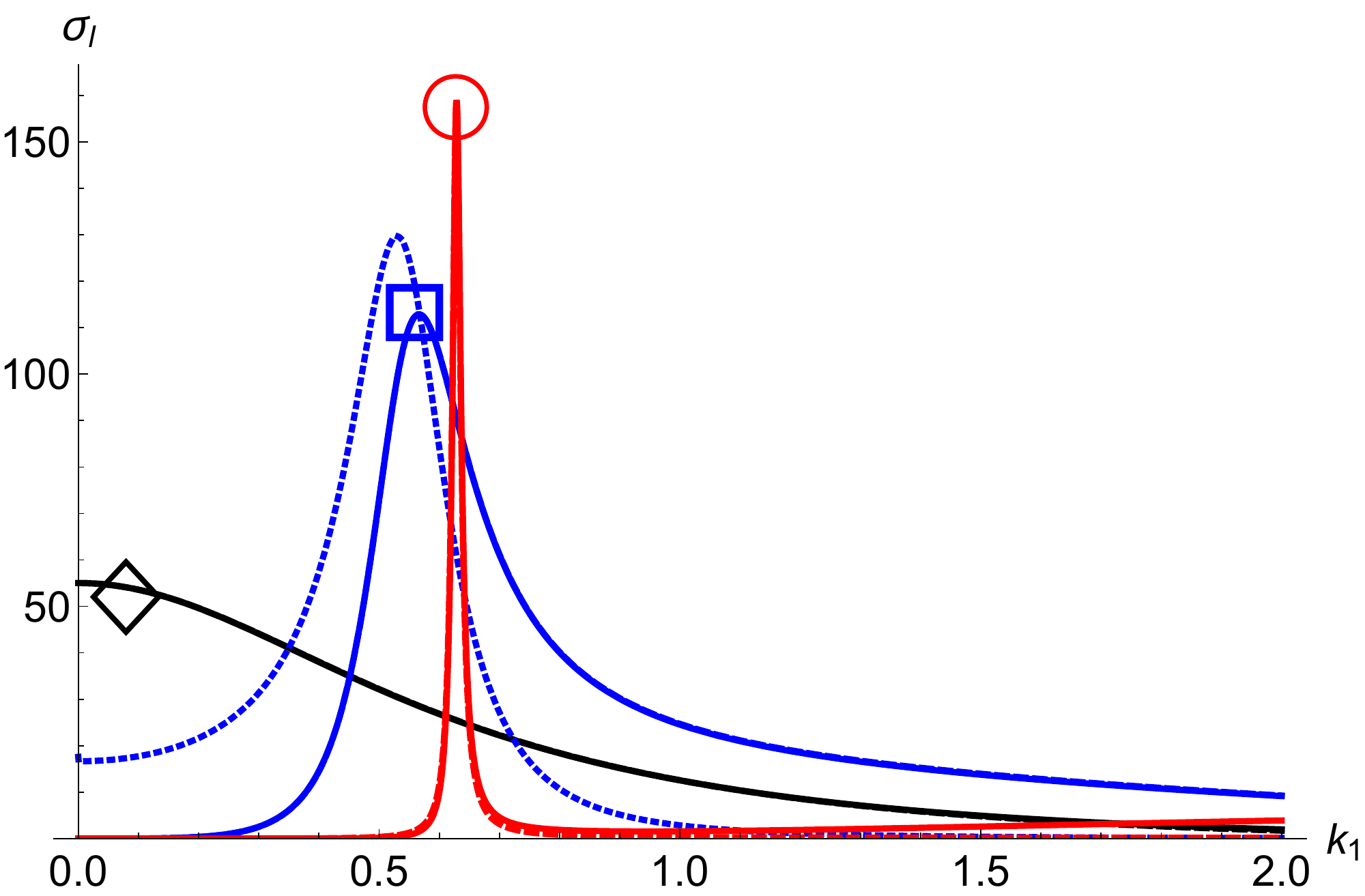}
\caption{(Color online) The exact partial cross section 
$\sigma^{(0)}_{l,11}(k_1)$ (solid) and 
the single-pole 
approximation (dashed) versus $k_1$ for
$l=0$ (black, $\diamond$), $l=1$ (blue, $\Box$) and $l=2$ (red, $\circ$) partial waves are almost indistinguishable.
Using only the Breit-Wigner approximation 
$\sigma_{l=1,11}^{(0)p,r}(k_1)$ (dotted, blue) of \eq{eq:BreitWigner} leads to significant 
deviations in shape and resonance position $k_r=\sqrt{E_r}=0.530$ (vertical black line).}
\label{fig:sigma_elastic}
\end{figure}

To be specific, we will consider the $s$-potential well-depth 
in Table~\ref{fig:Jost_l0}, which only supports one 
weakly bound state. It emerges as a zero of the Jost-function  
$f^{(0)}_{0,11}(k_1)=0$ at $k_1=i\mathcal{K}_1$ in the complex 
$k_1$-plane. 
Its zero contours are shown in Fig.~\ref{fig:Jost_l0}.
For wave-numbers close to threshold,  we assume that the Weierstrass expansion \eq{eq:Jost_Weierstrass_decomp}
is dominated by this value. 
All other zeros contribute to the background scattering phase $\eta_0^{p,b}$. 
\begin{figure}[H]
\centering\includegraphics[clip,width=\columnwidth]{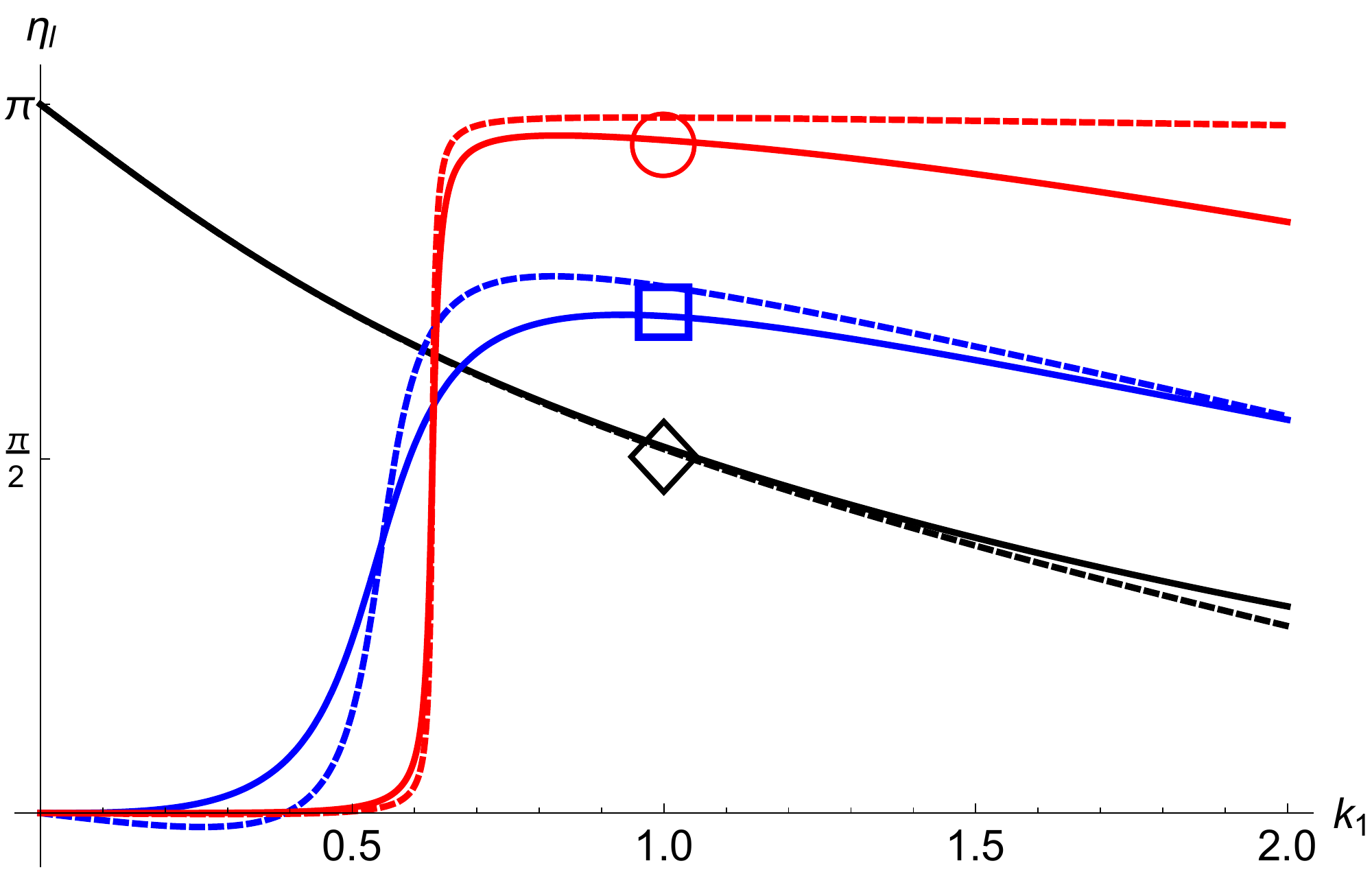}
\caption{(Color online) Exact scattering phase $\eta_l(k_1)$ (solid) and single-pole approximation $\eta_l^p(k_1)$ (dashed) versus $k_1$ for the 
 $l=0$ (black, $\diamond$), $l=1$ (blue, $\Box$) and $l=2$ (red, $\circ$) partial waves.}
\label{fig:scphases}
\end{figure}
Now , the single-pole expansion of the $S$-matrix reads
\begin{align}
	s^{(0)p}_{0,11}(k_1) &= s^{p,b}_{0,11}(k_1) s^{p,r}_{0,11}(k_1)
	= e^{2i\eta_0
^{p,b}} \frac{\mathcal{K}_1-ik_1}{\mathcal{K}_1+ik_1},
	\label{eq:sl0_pole}
\end{align}
and the background scattering phase is approximated from 
a Taylor series at $k_1=0$ of Eqs.~(\ref{eq:sl0}) and (\ref{eq:sl0_pole}) 
up to linear order as
\begin{align}
	\eta_0^{p,b}(k_1) &=
	\frac{1}{2i} 
	\log{\frac{s^{(0)}_{0,11}}{s_{0,11}^{p,r}}}
	\approx n \pi+\left(1-a_\text{sc} \mathcal{K}_1\right)\frac{k_1}{\mathcal{K}_1}.
	\label{eq:eta_pole_l0_linear}
\end{align}
The mathematical phase ambiguity is resolved physically by Levinson's 
theorem, counting the number of bound 
states $n\in \mathds{N}_0$ 
\footnote{For simplicity, we want to consider an integer 
 number of bound states and do not want to discuss the well known limit of 
half-bound states}. Moreover, we denote the $s$-wave scattering length 
  for the square-well potential \cite{WA2010} as
$a_\text{sc}=1-\tan\kappa_1/\kappa_1$.
The resonance contribution $\eta_0^{p,r}$  can be obtained from the second 
term in \eq{eq:sl0_pole}. For real $k_1$, it can be transformed into 
\begin{align}
\eta_0^{p,r}(k_1)&=-\arctan{\frac{k_1}{\mathcal{K}_1}},
\end{align}
using trigonometric relations \footnote{For a real angle $\alpha$, following 
useful identity has been employed repeatedly:   
$\arctan{\alpha}=\frac{1}{2i}\log{\frac{1+i \alpha}{1-i\alpha}}+n\pi$.}.
Consequently, the single-pole expansion of the total scattering 
phase \eq{eq:eta_decomp} is 
\begin{align}
	\eta_{0}^{p}(k_1) &=n\pi + 
	\left(1-a_\text{sc} \mathcal{K}_1\right)\frac{k_1}{\mathcal{K}_1}
	-\arctan{\frac{k_1}{\mathcal{K}_1}}.
\end{align}
In Fig.~\ref{fig:scphases}, the scattering phase $\eta_0^p$ of the pole 
expansion 
is compared to the exact scattering phase $\eta_0$.
It can be seen that they coincide almost perfectly for small wave-numbers.
We have chosen the potential depth parameter to accomodate one bound state, thus the scattering phase approaches  $\pi$ at threshold.
Analogously, we depict in Fig.~\ref{fig:sigma_elastic} the exact 
$s$-wave elastic cross sections of \eq{eq:sigma_partial_elastic} with the pole 
expansion. On the scale shown in the figure, there is hardly any 
difference.

\subsection{Scattering in the $p$-channel}
\label{subsubsec:sc_pwave}
In contrast to the $s$-channel, one can not have any virtual states for higher 
angular momenta. 
Resonant states with energies below the angular momentum barrier 
are called quasi-bound states.

\paragraph{Exact solution}

For $l=1$, the Jost function of \eq{eq:f_ii_squarewell} becomes
\begin{equation}
	\mathcal{f}^{(0)}_{1,11}(k_1) = 
	\frac{e^{ik_1}\sin q_1}{q_1^3}
	\left( k_1^2q_1 \cot q_1 +\kappa_1^2-i k_1 q_1^2  \right),
	\label{eq:Jost_l1}
\end{equation}
and the uncoupled $S$-matrix element can be derived 
 from \eq{eq:s_jost_uncoupled} as
\begin{equation}
	s^{(0)}_{1,11}(k_1) = e^{-2ik_1} \frac{k_1^2q_1 \cot q_1 +\kappa_1^2+ik_1q_1^2}{k_1^2q_1 \cot q_1 +\kappa_1^2-ik_1q_1^2}.
	\label{eq:sl1}
\end{equation}
The corresponding scattering phase is given by
\begin{equation}
	\eta_1(k_1) = n\pi -k_1 + \arctan{\frac{k_1q_1^2}{k_1^2q_1\cot q_1+\kappa_1^2}}.
\end{equation}
Again, the physical scattering phase is determined by the number of bound states $n$ in the $p$-wave potential.

\paragraph{Two-pole expansion}

\begin{figure}[t]
\centering\includegraphics[clip,width=\columnwidth]{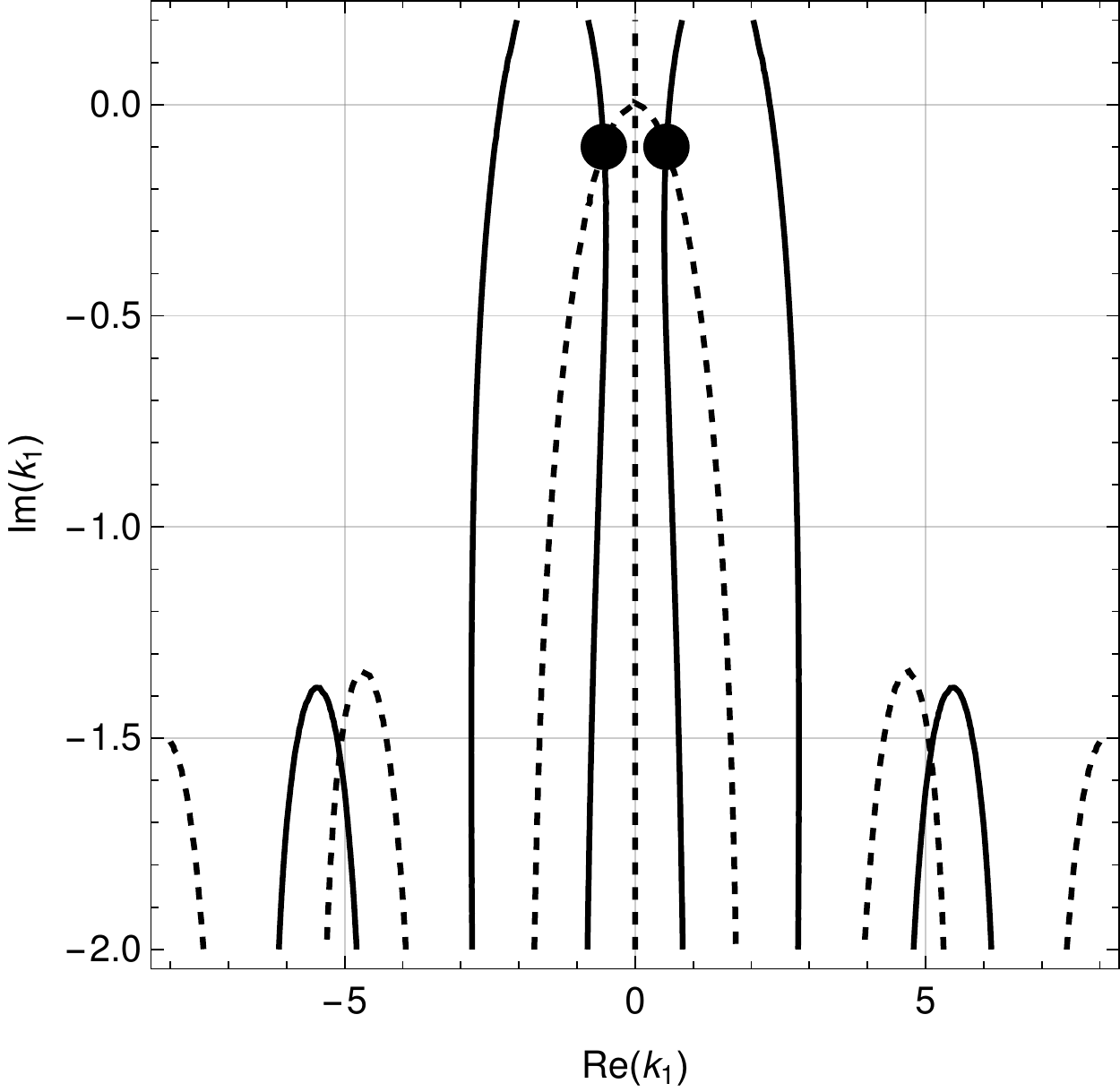}
\caption{Zero contour lines of the $p$-wave Jost function 
$\Re \mathcal{f}^{(0)}_{1,11}(k_1) = 0$ 
(solid) and 
$\Im \mathcal{f}^{(0)}_{1,11}(k_1) = 0$ (dashed) in the complex $k_1$-plane 
for a potential depth $\kappa_1^2=9$. The black dots indicate the positions 
$\mathcal{k}_1=0.539-0.100i$ and $-\mathcal{k}_1^*$.}
\label{fig:Jost_l1}
\end{figure}

The $p$-wave potential of Table~\ref{fig:Jost_l0}, 
supports one quasi-bound state. 
Mathematically, it is represented by
two closely spaced zeros $k_1=\{\mathcal{k}_1,-\mathcal{k}_1^*\}$
of the $p$-wave Jost function in the complex $k_1$-plane. 
Its zero contours are shown in Fig.~\ref{fig:Jost_l1}.
For low values of the wave-number, the Weierstrass expansion of the 
Jost-function is dominated by the pair of zeros. 
All the other zeros contribute cumulatively  
to the background scattering phase $\eta_1^{p,b}$. 
Then, the two-pole expansion of the $S$-matrix reads
\begin{align}
	s^{(0)p}_{1,11}(k_1) & = s^{p,b}_{1,11}\, s^{p,r}_{1,11}
	= e^{2i \eta^{p,b}_1}
	\frac{(\mathcal{k}_{1}+k_1)(\mathcal{k}_{1}^\ast - k_1)}{
	(\mathcal{k}_{1}^\ast + k_1)(\mathcal{k}_{1}-k_1)},
	\label{eq:sl1_pole}
\end{align}
and $\eta_1^{p,b}$ is found from Eqs.~(\ref{eq:sl1}) and (\ref{eq:sl1_pole})
\begin{equation}
	\label{eq:eta_pole_l1_linear}
	\eta_1^{p,b}(k_1) = \frac{1}{2i} 
	\log{\frac{s^{(0)}_{1,11}}{s^{p,r}_{1,11}}}
\approx 
n\pi +\frac{2 k_1 \Im{\mathcal{k}_1}}{|\mathcal{k}_1|^2},
\end{equation}
with a Taylor series at threshold $k_1=0$ up to linear order. 
The resonance scattering phase is defined by the second term in \eq{eq:sl1_pole}
and one finds for real $k_1$
\begin{align}
\eta_1^{p,r}(k_1)&= 
-\arctan{\frac{2 k_1 \Im{\mathcal{k}_{1}}}{|\mathcal{k}_{1}|^2-k_1^2}}.
\end{align}
Then, the total 
scattering phase of the pole expansion reads
\begin{align}
	\eta_1^p(k_1)&=
	n\pi+\frac{2 k_1\Im{\mathcal{k}_1}}{|\mathcal{k}_1|^2}
	-
	\arctan{\frac{2 k_1 \Im{\mathcal{k}_{1}}}{|\mathcal{k}_{1}|^2-k_1^2}}.
	\label{eta1}
\end{align}
In Fig.~\ref{fig:scphases}, the scattering phase $\eta_1^p$ of the pole 
expansion 
is compared to the exact solution $\eta_1$. The phase vanishes at zero energy 
as there is no bound state. While there is good qualitative overall 
agreement, there are noticeable deviations around the resonance position.

The resonant part of the elastic cross section can be evaluated by inserting 
$s_{1,11}^{p,r}$ in \eq{eq:sigma_partial_elastic} and one obtains the 
\emph{Breit-Wigner formula}~\cite{NEW2002}
\begin{align}
	\sigma^{(0)p,r}_{1,11} (k_1) &
	=\frac{12 \pi}{\Re^2 \mathcal{k}_1}
	\frac{(\Gamma_1/2)^2}{
	(k_1^2-\varkappa_1^2)^2+(\Gamma_1/2)^2},
	\label{eq:BreitWigner}\\
\Gamma_l & 
=4 \Re \mathcal{k}_l \Im \mathcal{k}_l, \\
E_r&=\varkappa_l^2  =\Re^2 \mathcal{k}_l - \Im^2 \mathcal{k}_l,
\end{align}
where we have introduced a resonance energy $E_r$ and a width 
$\Gamma_l$.
In Fig.~\ref{fig:sigma_elastic}, we compare the $p$-wave elastic cross sections 
\eq{eq:sigma_partial_elastic} 
from the exact $S$-matrix $s^{(0)}_{1,11}$ 
with the pole expansion $s^{(0)p}_{1,11}$ and find excellent agreement.
Using only the Breit-Wigner approximation  
\eq{eq:BreitWigner} leads to significant deviations.

\subsection{Scattering in the $d$-channel}
\label{subsubsec:sc_dwave}

$D$-wave scattering is qualitatively similar to the $p$-wave results. However, 
the increasing complexity of the exact solution conceals the physics. Only the 
pole expansion unveils the essential resonance features. 

\paragraph{Exact solution}

For $l=2$, the Jost function of \eq{eq:f_ii_squarewell} reads
\begin{gather}
	\mathcal{f}^{(0)}_{2,11}(k_1) =  
	\frac{e^{ik_1} \sin q_1}{q_1^5} 
	\left[ \left(k_1^4-3\kappa_1^2+3ik_1\kappa_1^2+
	k_1^2\kappa_1^2\right)  \right. \notag \\
	 \left. \times q_1  \cot q_1 - 
	i\left(k_1^5+3i\kappa_1^2+3k_1\kappa_1^2+k_1^3\kappa_1^2\right)
	 \right].
	\label{eq:Jost_l2}
\end{gather}
The $S$-matrix element follows from \eq{eq:s_jost_uncoupled} as
\begin{align}
	s^{(0)}_{2,11}(k_1) &= \frac{\mathcal{f}^{(0)}_{2,11}(-k_1)}{
	\mathcal{f}^{(0)}_{2,11}(k_1)},
	\label{eq:sl2}
\end{align}
and for real $k_1$ the scattering phase reads
\begin{align}
	& \eta_2(k_1)  = n\pi - k_1 \notag \\ & -\arctan 
	\frac{k_1[(k_1^4+3\kappa_1^2+k_1^2\kappa_1^2)-3\kappa_1^2 q_1 \cot q_1]}{
	(k_1^4-3\kappa_1^2+k_1^2\kappa_1^2)q_1 \cot q_1+3\kappa_1^2}.
\end{align}

\begin{figure}[t]
\centering
\includegraphics[clip,width=\columnwidth]{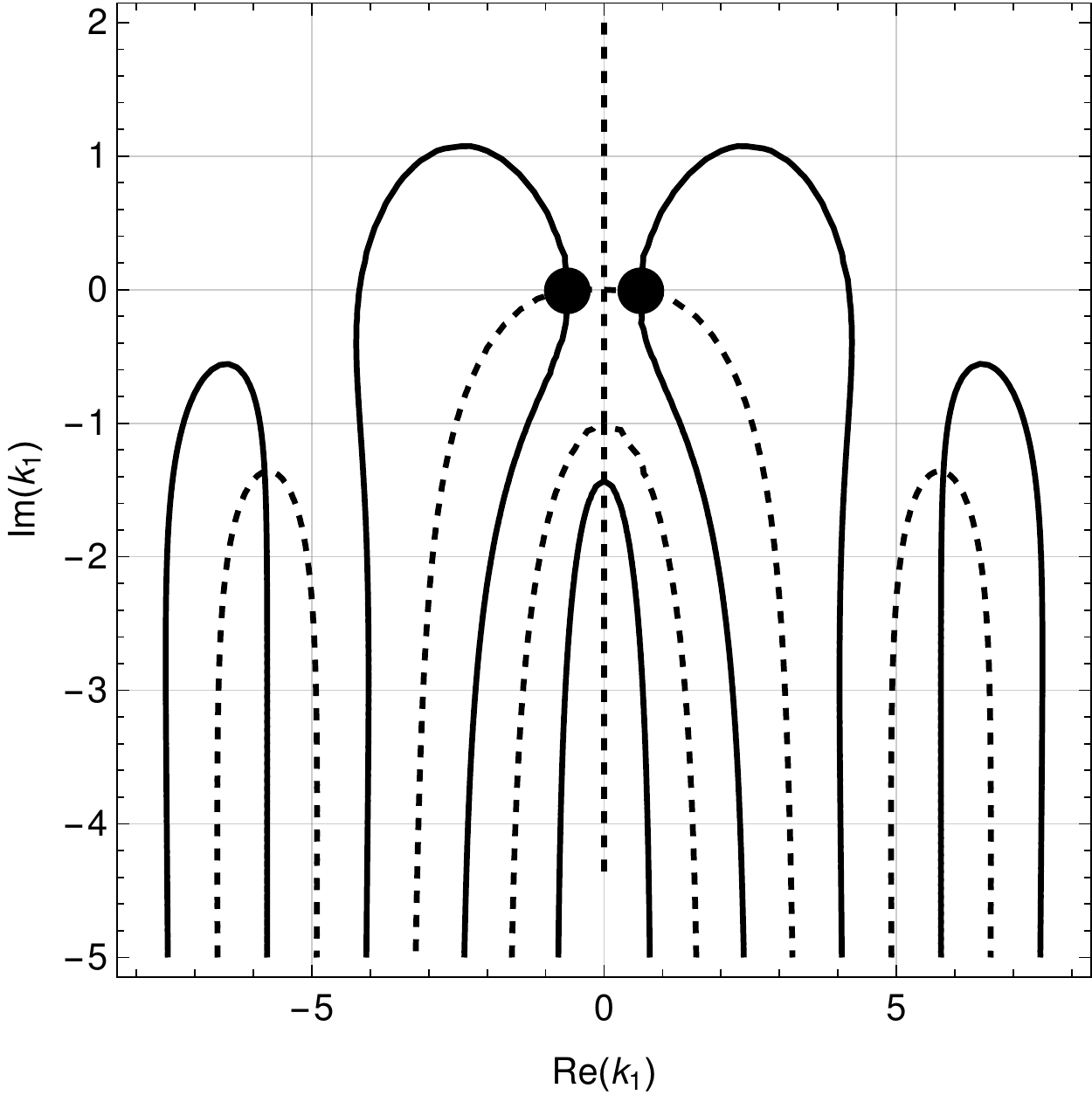}
\caption{Zero contour lines of the $d$-wave Jost function 
$\Re \mathcal{f}^{(0)}_{2,11}(k_1) = 0$ 
(solid) and 
$\Im \mathcal{f}^{(0)}_{2,11}(k_1) = 0$ (dashed) in the complex $k_1$ plane for a potential depth $\kappa_1^2=19.5$. 
The black dots indicate the positions $\mathcal{k}_2=0.628-0.009i$ 
and $-\mathcal{k}_2^*$.}
\label{fig:Jost_l2}
\end{figure}

\paragraph{Two-pole expansion}

The quasi-bound state of the $d$-wave potential emerges from two zeros 
 $k_1=\{\mathcal{k}_2, -\mathcal{k}_2^*\}$ of the 
Jost function $f^{(0)}_{2,11}(k_1)$ in the complex $k_1$-plane. 
Its zero contours are shown in Fig.~\ref{fig:Jost_l2}.
We assume that the pole expansion of the $S$-matrix 
\begin{align}
	s^{(0)p}_{2,11}(k_1) & = 
	s^{p,b}_{2,11}(k_1)\cdot s^{p,r}_{2,11}(k_1) \notag \\
	& = e^{2i \eta^{p,\text{bg}}_2(k_1)}
		\frac{(\mathcal{k}_2+k_1)(\mathcal{k}_2^\ast - k_1)}{
	(\mathcal{k}_2^\ast + k_1)(\mathcal{k}_2-k_1)},
	\label{eq:sl2_pole}
\end{align}
is dominated 
by the two poles for low $k_1$. 
The effect of all the other poles contributes to the 
background scattering phase $\eta_2^{p,b}$.
Using  Eqs.~\eqref{eq:sl2} and 
\eqref{eq:sl2_pole}, the background scattering phase is found from a Taylor series at $k_1=0$ up to linear order as
\begin{equation}
	\eta^{p,b}_2(k_1) = \frac{1}{2i} \log 
	\frac{s^{(0)}_{2,11}}{s^{p,r}_{2,11}}
	\approx
	n\pi + 
	\frac{2 k_1 \Im \mathcal{k}_2}{|\mathcal{k}_2|^2}.
	\label{eq:eta_pole_l2_linear}
\end{equation}
It has the same structure as the $p$-wave result \eqref{eq:eta_pole_l1_linear}, which presumably holds also for higher angular momenta. It holds definitively for the resonance phase as the structure of the Weierstrass expansion is identical.
Therefore, the total scattering phase for the pole expansion reads
\begin{equation}
	\eta_2^p(k_1) = 
	n\pi + 
	\frac{2 k_1 \Im \mathcal{k}_2}{|\mathcal{k}_2|^2} 
	- \arctan \frac{2k_1 \Im \mathcal{k}_2}{|\mathcal{k}_2|^2-k_1^2},
\end{equation}
analogously to \eq{eta1}.

In Fig.~\ref{fig:scphases}, the scattering phase $\eta_2^p$ of the pole expansion 
is compared to the exact solution $\eta_2$. Around the resonance position and for larger $k_1$, $\eta_2^p$ deviates from the exact solution.
In Fig.~\ref{fig:sigma_elastic}, the $d$-wave elastic cross sections \eq{eq:sigma_partial_elastic} found from the exact $S$-matrix $s^{(0)}_{2,11}$ and found from the pole expansion $s^{(0)p}_{2,11}$ are compared. Only for larger $k_1$ the solutions start to deviate from each other, the resonant behavior is explained perfectly by the pole expansion.

\section{Two-body loss rates}
\label{sec:two_body}

\begin{figure}[t]
\centering\includegraphics[clip,width=\columnwidth]{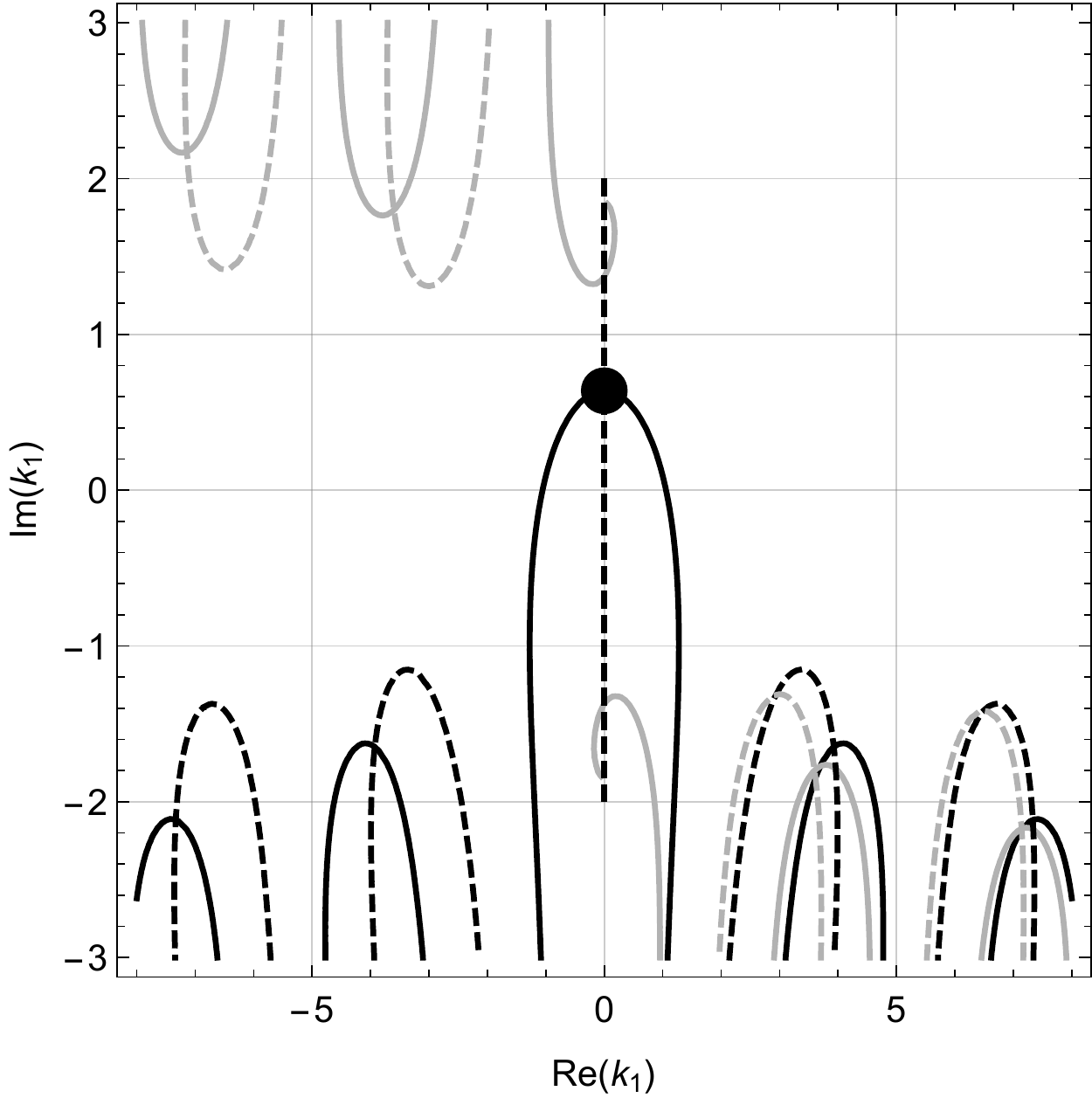}
\caption{Zero contours lines of the $s$-wave Jost functions in channel 1 and 2:  
$\Re \mathcal{f}^{(0)}_{0,11}(k_1)=0$ (solid, black), 
$\Im \mathcal{f}^{(0)}_{0,11}(k_1)=0$ (dashed, black) and 
$\Re \mathcal{f}^{(0)}_{0,22}(k_1)=0$ (solid, gray), 
$\Im \mathcal{f}^{(0)}_{0,22}(k_1)=0$ (dashed, gray) in the complex $k_1$-
plane. The black dot indicates the position $k_1= i\mathcal{K}_1=0.638 i$. The potential parameters are given in Table~\ref{tab:simparam}.}
\label{fig:Jost_tc_l0}
\end{figure}

\begin{figure}[ht]
\centering\includegraphics[clip,width=\columnwidth]{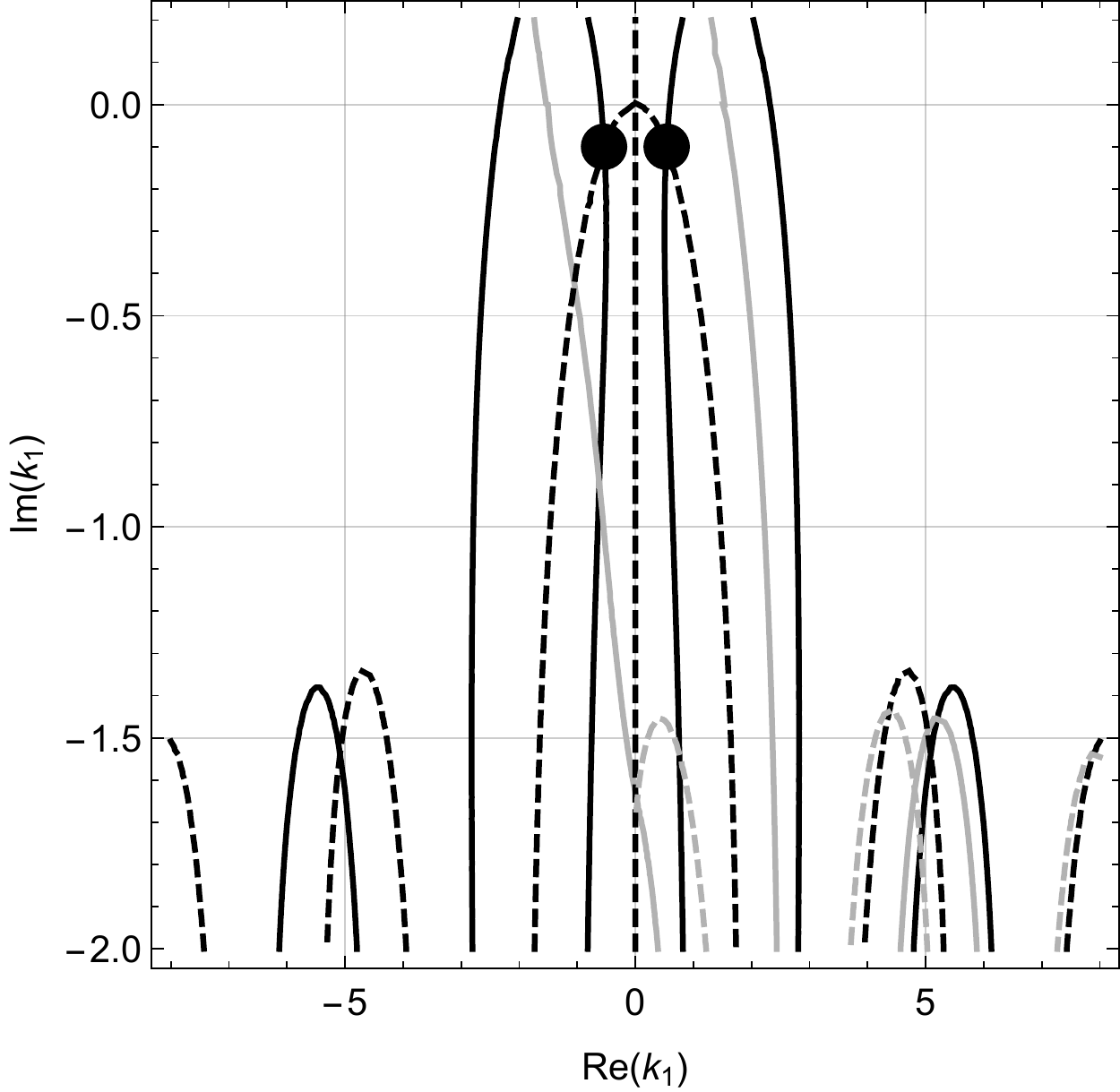}
\caption{Zero contour lines of $p$-wave Jost function in channels 1 and 2: 
$\Re \mathcal{f}^{(0)}_{1,11}=0$ (solid, black), 
$\Im \mathcal{f}^{(0)}_{1,11}=0$ (dashed, black) and 
$\Re \mathcal{f}^{(0)}_{1,22}=0$ (solid, gray), $\Im 
\mathcal{f}^{(0)}_{1,22}=0$ (dashed, gray) in the complex $k_1$ plane. The 
black dots indicate the positions $\mathcal{k}_1=0.539-0.100i$
and $-\mathcal{k}^*_1$. The 
potential parameters are given in Table~\ref{tab:simparam}.}
\label{fig:Jost_tc_l1}
\end{figure}

Now, we will extent the discussion from the single-channel case to the 
coupled two-channel case within the DWBA.
We use the same potential parameters for the upper channel as in 
the single channel case given in Table~\ref{tab:param_single}.
The weakly bound $s$-state in the upper channel as well as the quasi-bound 
$p$-  and $d$-states will dominate the behavior of the partial two-body loss 
rates as the channels are only weakly coupled to the loss channels.
Given a sufficiently large separation of channel threshold energies 
$\Delta_{11}\gg \Delta_{22}$, one can assume that the potential 
depth $\kappa_2^2$ 
of channel $2$ does not influence the scattering behavior of channel $1$.
For simplicity, we have chosen them equal. All other parameters are listed in 
Table~\ref{tab:simparam}.

\begin{table}[th]
\centering
\begin{ruledtabular}
\begin{tabular}{lcrrr}
Angular momentum & $l $& $0$ & $1$ & $2$ \\
\hline
Potential depth & $\kappa_2^2$ & $4.0$ & $9.0$ & $19.5$ \\
Threshold energy channel 1 & $\Delta_{11}$ & $0.0$ & $0.0$ & $0.0$ \\
Threshold energy channel 2 &$\Delta_{22}$ & $-3.0$ & $-3.0$ & $-3.0$ \\
Coupling strength &$v_{12}$ & $0.1$ & $0.1$ & $0.1$
\end{tabular}
\end{ruledtabular}
\caption{Potential parameters of two-channel scattering used in the examples.}
\label{tab:simparam}
\end{table}

The transition amplitude  between the channels is determined by the $S$-matrix 
element $s^{(1)}_{l,12}$ of \eq{eq:s12_DWBA_squarewell}. 
Its  poles follow from the zeros of the uncoupled Jost functions 
$\mathcal{f}^{(0)}_{l,11}$, $\mathcal{f}^{(0)}_{l,22}$ of channel $1$ and $2$, respectively. 

We present the zero contours of the $s$-wave Jost functions $\mathcal{f}^{(0)}_{l,11}$ and $\mathcal{f}^{(0)}_{l,22}$
in Fig.~\ref{fig:Jost_tc_l0}, and in 
Fig.~\ref{fig:Jost_tc_l1} the $p$-wave result. 
It can be seen that the zeros of $\mathcal{f}^{(0)}_{l,22}$ are further away 
from the origin than the smallest zero(s) of $\mathcal{f}^{(0)}_{l,11}$ marked 
in blue. A similar picture arises for $l=2$.
Therefore, we assume that the pole expansion of $s_{l,12}^{(1)}$ is 
dominated by 
the zeros of $\mathcal{f}^{(0)}_{l,11}$ at 
$k_1 = i \mathcal{K}_1$ (l=0) and  the pair of zeros
$k_1=\{\mathcal{k}_l,-\mathcal{k}_l^\ast\}$ for $l=1,2$. 
Non-resonant features contribute to the background scattering phase 
$\eta_{l,12}^{p,b}$.  We obtain for the pole expansion
\begin{align}
	s^{(1)p}_{0,12}(k_1) = & 
	\frac{a_0 {k_1}^{\frac{1}{2}} 
	e^{\xi^{p,b}_{0,12}(k_1)}}{\mathcal{K}_1+ik_1}, 
	\label{subeq-1:s12_pole} \\
	s^{(1)p}_{l,12}(k_1) = & 
	\frac{a_l k_1^{\frac{2l+1}{2}} e^{\xi^{p,b}_{l,12}(k_1)}}{
	(\mathcal{k}_l^\ast+k_1)(\mathcal{k}_l-k_1)},
	\label{subeq-2:s12_pole}
\end{align}
where the second line holds for $l>0$ and the background attenuation 
coefficients 
are given by
\begin{align}
	\xi^{p,b}_{0,12}(k_1) = & \log \left[ s^{(1)}_{0,12}(k_1)
	\frac{ (\mathcal{K}_1+ik_1)}{a_0 \sqrt{k_1} } \right]\notag\\
	&=  \frac{1}{2}(b_0 k_1 + c_0 k_1^2 + \ldots), 
	\label{subeq-1:eta_bg_tcm} \\
	\xi^{p,b}_{l,12}(k_1) = &
	\log \left[ s^{(1)}_{l,12}(k_1)
	\frac{(\mathcal{k}_l^\ast+k_1)(\mathcal{k}_l-k_1)}{a_l k_1^{\frac{2l+1}{2}}
 } \right] \notag \\
	= & \frac{1}{2}\left(b_l k_1 + c_l k_1^2 + \ldots \right).
	\label{subeq-2:eta_bg_tcm}
\end{align}
Here, a Taylor expansion of the background scattering phases 
around $k_1=0$ yields expansion coefficients 
$a_l,b_l,c_l,\ldots$, which depend on the potential parameters and the 
pole positions. 

From this expansion, one can obtain the two-body loss rates of  
\eq{eq:partial_twobody}. In lowest 
order, they read
\begin{subequations}
	\begin{align}
	\beta^{(1)p}_0(k_1) = & \frac{2\pi |a_0|^2}{k_1^2+\mathcal{K}_1^2}, \label{
subeq-1:beta_k_l0} \\
	\beta^{(1)p}_l(k_1) = & \frac{2\pi (2l+1)|a_l|^2}{\left(k_1^2-\varkappa_l^2
\right)^2 +(\Gamma_l/2)^2} k_1^{2l}
 \label{subeq-2:beta_k_l},
	\end{align}
	\label{eq:betal_parametrized}
\end{subequations}
An important feature of the pole expansion of the two-body loss rates in the 
Eqs.~\eqref{eq:betal_parametrized} is the power law behavior $k_1^{2l}$. This is known as the \emph{Wigner threshold 
behavior}~\cite{JULIENNE1989,Bethe1935,WIGNER1948}.

\begin{figure}[ht]
\centering\includegraphics[clip,width=\columnwidth]{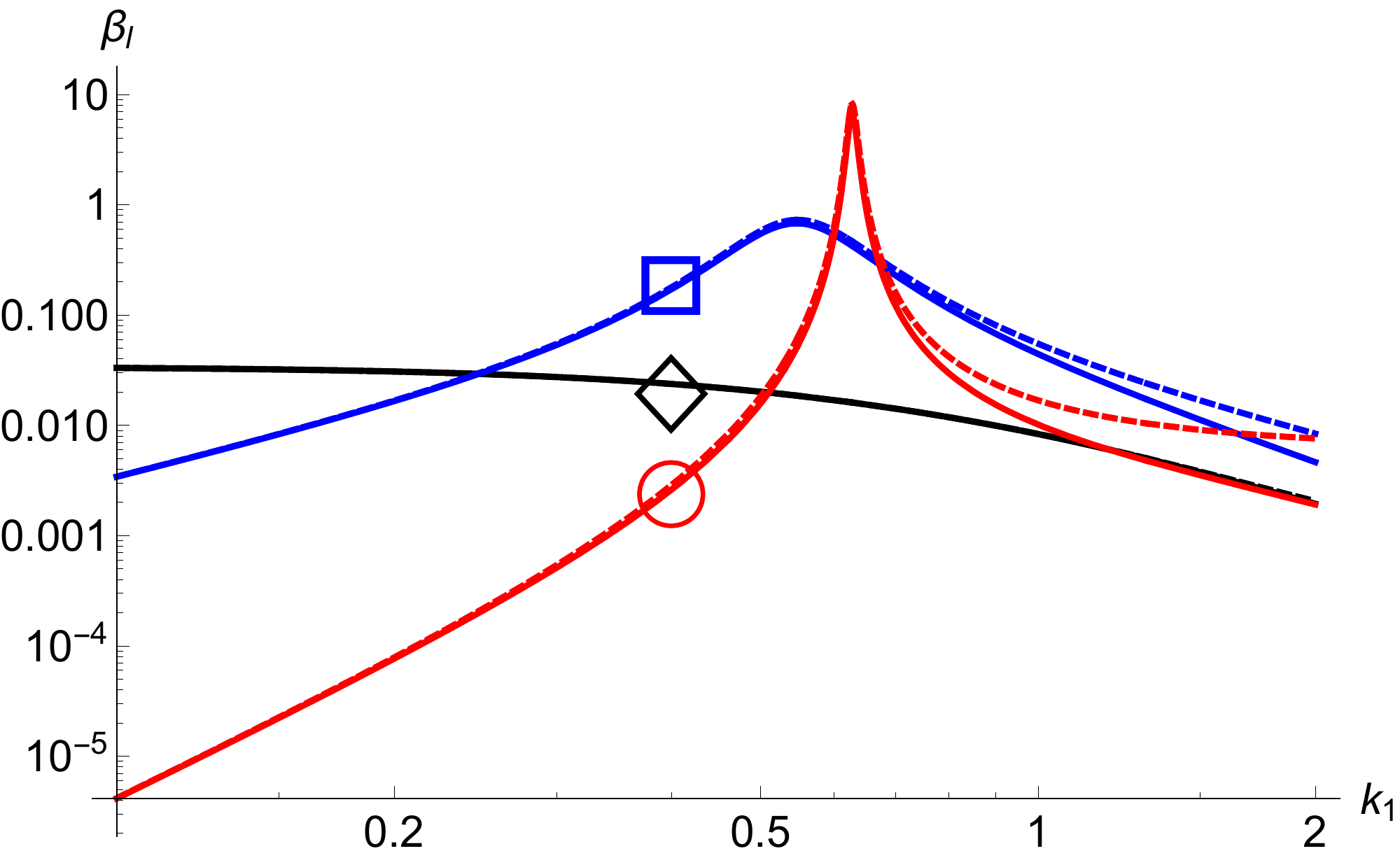}
\caption{(Color online) DWBA partial-wave two-body loss rates $\beta^{(1)}_l$ (solid) and 
the single-pole approximation $\beta^{(1)p}_l$ (dashed) versus channel wave-number 
$k_1$ for $l=0$ (black, $\diamond$), $l=1$ (blue, $\Box$), $l=2$ (red, $\circ$) for the potential parameters of Table~\ref{tab:simparam}.}
\label{fig:beta_l0l1l2}
\end{figure}

In Fig.~\ref{fig:beta_l0l1l2}, we compare the partial two-body loss rates of 
the DWBA \eq{eq:partial_twobody} with the pole expansion Eqs.~\ref{eq:betal_parametrized}. The pole expansion describes the low-energy and the resonant 
behavior of the two-body loss rates very well and starts to deviate from the 
DWBA solution only beyond the resonance positions. 
 
\section{Analysis of synthetic scattering data with shape functions}
\label{sec:effective_theory}

Here, we employ the pole expansion of the partial two-body loss rates 
\eq{eq:betal_parametrized} to analyze the synthetic scattering 
data of~\cite{COP2016} for heteronuclear $^{20}$Ne-$^{21}$Ne PI and AI collisions. In the simulations of~\cite{COP2016}, the two-body 
loss rate $\beta$ was calculated 
in terms of a partial-wave decomposition
\begin{equation}
	\beta = \frac{1}{2}\sum_l \beta_l,
	\label{eq:beta_betal}
\end{equation}
with the partial two-body loss rates $\beta_l$.The factor $1/2$ accounts for the collisions of non-identical particles~\cite{WEINER1999}. 
In the experimental temperature regime 
$T\simeq \unit{1}{\milli \kelvin}$, only few partial-waves 
$l\leq 2$ contribute to the total loss rate. 

In order to obtain a high quality interpolation of the two-body loss rates of  the available synthetic scattering data, 
we have to consider the higher order corrections
of Eqs.~(\ref{subeq-1:eta_bg_tcm},\ref{subeq-2:eta_bg_tcm})
in the two-body loss rates of \eq{eq:partial_twobody} and find 
\begin{subequations}
	\begin{align}
	\beta^{p}_0(k) = & \frac{2\pi |a_0|^2}{k^2+\mathcal{K}_1^2}e^{\Re c_0 k^2} \label{subeq-1:beta_k_l0_fit}, \\
	\beta^{p}_1(k) = & \frac{6\pi |a_1|^2}{
	\left(k^2-\varkappa_1^2\right)^2 + (\Gamma_1/2)^2} k^2 
	e^{\Re c_1 k^2}
	\label{subeq-2:beta_k_l1_fit}, \\
	\beta^{p}_2(k) = & \frac{10\pi |a_2|^2}{
	\left(k^2-\varkappa_2^2\right)^2 + (\Gamma_2/2)^2} k^4 
	e^{\Re (c_2 k^2+e_2 k^4)}
	\label{subeq-3:beta_k_l2_fit}.
	\end{align}
	\label{eq:betal_parametrized_fit}%
\end{subequations}
In the additional attenuation factors of the 
two-body loss rates all odd power of $k_1$ vanish as the coefficients $b_l, d_l, \ldots$ are purely imaginary.

A least square fit of the partial two-body loss rates in the Eqs.~\eqref{subeq-1:beta_k_l0_fit},~\eqref{subeq-2:beta_k_l1_fit},~\eqref{subeq-3:beta_k_l2_fit} to the synthetic data of Ref.~\cite{COP2016} leads to the coefficients given in Table~\ref{tab:coeffs}.
\begin{table}[th]
\centering
\begin{ruledtabular}
\begin{tabular}{crrr}
 & $l=0$ & $l=1$ & $l=2$ \\
\hline
$a_l$ & $0.1477$ & $0.1326$ & $2.3141$ \\
$\Re c_l$ & $57.3641$ & $76.8736$ & $-64.7843$ \\
$\Re e_l$ & --- & --- & $-642.3825$ \\
$\mathcal{K}_l$ & $0.0408$ & --- & --- \\
$\Re \mathcal{k}_l$ & --- & $1.5 e^{-10}$ & $0.0304$ \\
$\Im \mathcal{k}_l$ & --- & $-0.0198$ & $-0.0100$ \\
\end{tabular}
\end{ruledtabular}
\caption{Expansion coefficients and pole positions found from fit of $\beta^p_l$ to the synthetic data.}
\label{tab:coeffs}
\end{table}

\begin{figure}[ht]
\centering\includegraphics[clip,width=\columnwidth]{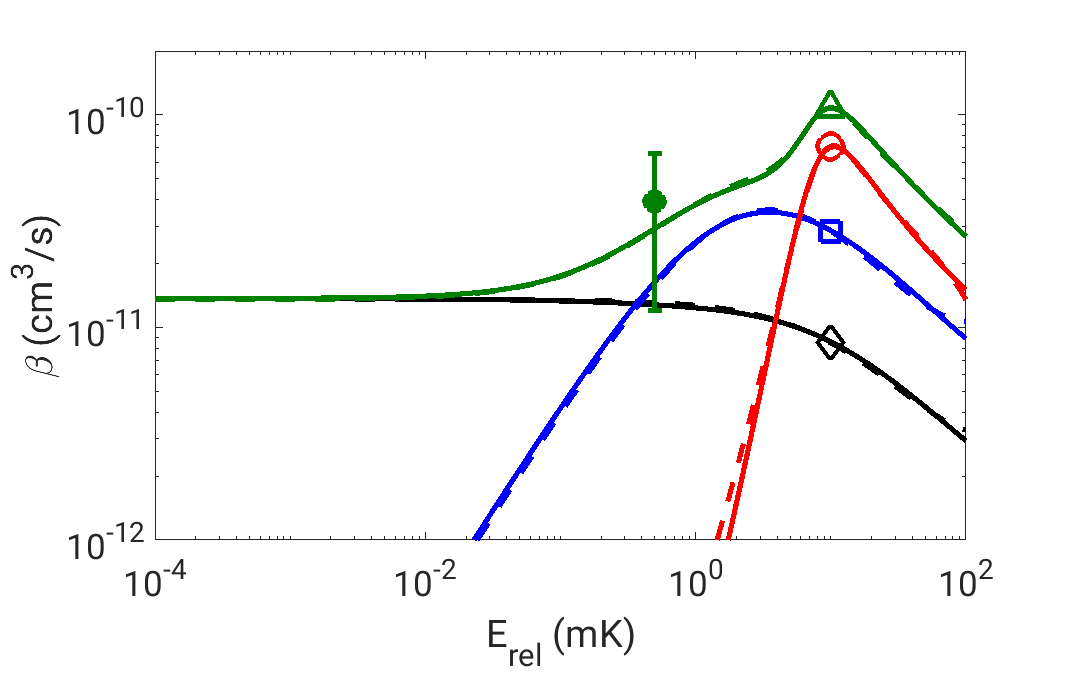}
\caption{(Color online) Synthetic scattering two-body loss rates $\beta^\text{synt}$ (solid, green,$\bigtriangleup$) and pole expansion $\beta^p$ (dashed, green,$\bigtriangleup$) with the coefficients of Table~\ref{tab:coeffs} versus relative collisions energy $E_\text{rel}=\hbar k^2/2\mu$ in units of $k_B$ for $^{21}$Ne-$^{20}$Ne collisions. Additionally shown are the partial-wave contributions $1/2\times \beta^\text{synt}_l$ [solid, (black, $\diamond$), (blue, $\Box$), (red, $\circ$)], $1/2\times \beta^p_l$ [dashed, (black, $\diamond$), (blue, $\Box$), (red, $\circ$)] for $l=0,1,2$, respectively, and experimental data point (green filled circle with error bars).}
\label{fig:fit_simulations}
\end{figure}

In Fig.~\ref{fig:fit_simulations}, we compare 
the synthetic data with the optimal fits from the shape functions obtained 
from the pole expansion.
We also included the single available experimental data point from 
Ref.~\cite{Schutz2012} to the picture. It can be seen that the pole 
expansion agrees very well with the 
synthetic data as well as with the experimental data point. 
Technically speaking, the two-body loss rates should be averaged thermally 
to match the experimental data point. However, the width of the 
relative velocity distribution in the thermal gas is large compared to the 
widths of the two-body loss rates. Moreover, the error bar for the 
experimental data point is large, too. Therefore, we neglect the effects of 
thermal averaging.

\section{Conclusion}
\label{sec:conclusion}

In this article, we have presented a coupled two-channel model for the
reactive collisions of atoms at low collision energies. Transitions from 
the elastic scattering channel to the lower ionization channel 
models loss of atoms in two-body collisions, as observed in auto-ionization
and Penning ionization processes. In 
particular, we study the 
two-channel square-well model. On one hand, 
this model can be solved in closed form and, on the other hand, one can use the pole approximation to obtain physically motivated shape functions from it. 
To extract useful approximations, we employed the 
distorted-wave Born approximation and studied the poles in the 
complex $k_1$ plane for the lowest $s$-, $p$- and $d$-partial-waves. 
From this analysis, we obtain simple analytic expressions for the 
partial two-body loss rates. Fitting these analytic 
two-body loss rates to available synthetic scattering 
data~\cite{COP2016} on cold 
heteronuclear $^{20}$Ne--$^{21}$Ne collisions gives very good agreement and 
also matches the experimental data point of Ref.~\cite{Schutz2012}.

\section*{Acknowledgements}

We are very grateful to E. Tiesinga, S. Kotochigova, P.~S. Julienne, and 
C. Williams for their hospitality at NIST and for inspiring discussions. 
In particular, we thank S. Kotochigova for providing the Ne* molecular 
potentials to us. Moreover, we thank A. Martin, J. Sch\"utz and G. Birkl for numerous helpful discussions and providing experimental data.
RW acknowledge gratefully travel support from the German Aeronautics and Space Administration (DLR) through grant 50 WM 1137.

\appendix
\section{Riccati-Bessel functions}
\label{sec:bessel}
Bessel function are a core element of three-dimensional scattering theory. In order to avoid definitional ambiguities, we  use 
the Riccati-Hankel functions 
\begin{equation}
	\hat{h}^\pm_l(z) \equiv \hat{n}_l(z) \pm i \hat{\jmath}_l(z),
\end{equation}
as in Ref~\cite{TAY2006},
where $\hat{\jmath}_l(z)$ is the Riccati-Bessel function and $\hat{n}_l(z)$ is the Riccati-Neumann function \cite{ABRAMOWITZ1965}
\begin{align}
	\hat{\jmath}_l(z) \equiv & \sqrt{\frac{\pi z}{2}} J_{l+\tfrac{1}{2}}(z), \\
	\hat{n}_l(z) \equiv & (-1)^l \sqrt{\frac{\pi z}{2}} J_{-l-\tfrac{1}{2}}(z),
\end{align}
with the Bessel function of the first kind $J_{l}(z)$.
The Riccati-Hankel functions have the symmetry properties
\begin{equation}
	\hat{h}^\pm_l(-z) = (-1)^{l+1} \hat{h}^\mp_l(z).
\end{equation}
The behavior of $\hat{\jmath}_l(z)$ for small arguments $z$ is given by
\begin{equation}
	\hat{\jmath}_l(z) \stackrel[z\to 0]{}{\longrightarrow} \frac{z^{l+1}}{(2l+1)!!}.
	\label{eq:RiccatiBessel_zero}
\end{equation}

\section{Analytic pole approximation}
One can determine the zeros of the Jost function $\mathcal{f}^{(0)}_{l,11}$
 in the complex $k_1$ plane  either numerically or analytically by 
introducing simple approximations. 
In the case of the $s$-wave Jost function \eq{eq:Jost_l0}, one can use complex transformations
$k_1 \equiv z\kappa_1$, $z\equiv -i \cos w$,
to find
\begin{equation}
	\kappa_1 \sin w = w+n\pi,\quad n \in \mathds{N}_0^+.
	\label{eq:sc_l0_approx_pole}
\end{equation}
By solving this equation for $w$ and for all $n$, we obtain all zeros of 
$\mathcal{f}^{(0)}_{0,11}$. For the zeros close to the origin of the complex plane, we can assume 
$|k_1|/\kappa_1 \ll 1$ and it follows $w \simeq \pi/2$. A Taylor series of 
\eqref{eq:sc_l0_approx_pole} at $w=\pi/2$ to second order leads to
  \begin{equation}
	w_\pm = \frac{-1+\pi \kappa_1\pm \sqrt{1-\pi \kappa_1-2n\pi\kappa_1+2 
\kappa_1^2}}{\kappa_1}.
	\label{eq:Jost_l0_zeros_approx}
\end{equation}
The zero of the Jost function due to the first bound state of the potential 
is given by $w_+$ and $n=0$, the zero of the Jost function due to the second 
bound state by $w_+$ and $n=1$ and so forth. For $N$ bound states present 
in the scattering potential, the zero of the Jost function of the first 
virtual state is given by $w_+$ for $n=N$, for the second virtual state by $w
_+$ for $n=N+1$ and so forth up to $n=N+N'-1$ for $N'$ virtual states 
present. The zeros of the Jost function due to scattering resonances are 
given by the two solutions $w_\pm$ for $n>N+N'-1$.

For $\kappa_1=4$ and $n=0$, we find $w_+=1.895$ from \eq{eq:Jost_l0_zeros_approx}. Resubstitution leads to a zero of the Jost function at $\mathcal{k}_1 = i\mathcal{K}_1$ with $\mathcal{K}_1 = 0.637$. The numerically determined value is given by $\mathcal{K}_1 = 0.638$. 

\bibliographystyle{apsrev4-1}
\bibliography{squarewell} 
\end{document}